\documentclass[a4paper]{article}
\usepackage[margin=1in]{geometry}

\newcommand{\xdna}{x_{(j,\bm{n})}}
\newcommand{\LM}[1]{\bm{L}^{\text{#1}}}
\newcommand{\LMb}[1]{\bm{L}^{\text{#1}}}
\newcommand{\LMbn}[2]{\bm{L}^{\text{\~{#1},\~{#2}}}}
\newcommand{\x}[1]{\bm{x}^{\text{#1}}}
\newcommand{\dx}[1]{\bm{\delta x}^\text{#1T}}

\newcommand{\ep}{\bm{e}_p}

\newcommand{\bx}{\bm{x}}
\newcommand{\bL}{\bm{L}}
\newcommand{\bR}{\bm{R}}
\renewcommand{\l}{\mathrm{l}}
\renewcommand{\a}{\mathrm{a}}

\renewcommand{\aa}{\mathrm{aa}}
\newcommand{\T}{^\mathrm{T}}

\newcommand{\eff}{_{\mathrm{eff}}}

\usepackage{amsmath,amsthm,amssymb,graphicx,bm,authblk,breakcites}

\author[1,2]{Edgar Herrera-Delgado}
\author[1,3]{Ruben Perez-Carrasco}
\author[1]{James Briscoe*}
\author[2]{Peter Sollich*}

\affil[1]{The Francis Crick Institute, 1 Midland Road, London, NW1 1AT, UK}
\affil[2]{Department of Mathematics, King's College London, Strand, London WC2R 2LS, UK}
\affil[3]{Department of Mathematics, University College London, Gower Street, London WC1E 6BT, UK}

\date{}

\title{Memory functions reveal structural properties of gene regulatory networks}

\begin{document}
\maketitle
\noindent* james.briscoe@crick.ac.uk\\
* peter.sollich@kcl.ac.uk

\begin{abstract}
Gene regulatory networks (GRNs) control cellular function and decision making during tissue development and homeostasis. Mathematical tools based on dynamical systems theory are often used to model these networks, but the size and complexity of these models mean that their behaviour is not always intuitive and the underlying mechanisms can be difficult to decipher. For this reason, methods that simplify and aid exploration of complex networks are necessary. To this end we develop a broadly applicable form of the Zwanzig-Mori projection. By first converting a thermodynamic state ensemble model of gene regulation into mass action reactions we derive a general method that produces a set of time evolution equations for a subset of components of a network. The influence of the rest of the network, the bulk, is captured by memory functions that describe how the subnetwork reacts to its own past state via components in the bulk. These memory functions provide probes of near-steady state dynamics, revealing information not easily accessible otherwise. We illustrate the method on a simple cross-repressive transcriptional motif to show that memory functions not only simplify the analysis of the subnetwork but also have a natural interpretation. We then apply the approach to a GRN from the vertebrate neural tube, a well characterised developmental transcriptional network composed of four interacting transcription factors. The memory functions reveal the function of specific links within the neural tube network and identify features of the regulatory structure that specifically increase the robustness of the network to initial conditions. Taken together, the study provides evidence that Zwanzig-Mori projections offer powerful and effective tools for simplifying and exploring the behaviour of GRNs.
\end{abstract}

\section{Introduction}
\indent Biological systems are complex, comprising multiple interacting components. In many cases this complexity makes it difficult to identify underlying mechanisms and to understand the function of a system. Gene regulatory networks (GRNs) are an example of this problem \cite{Levine2005}. A GRN comprises the set of interacting genes responsible for the development, differentiation or homoeostasis of a tissue and provides a formal system-level, causative explanation for gene regulation. In physical terms, a GRN consists of modular DNA sequences -- cis regulatory elements -- that bind to specific sets of transcriptional activators and repressors, which control the expression of associated genes. Some of the regulated genes are themselves transcriptional regulators. Thus, at the core of a GRN is a recursive set of regulatory links that forms a transcriptional network, the dynamics of which is responsible for the spatial and temporal patterns of gene expression.

Attempts have been made to map large transcriptional networks, yet even for relatively small networks, the number of links and the feedback within the system make intuitive understanding difficult to obtain. Various computational models have been developed to address this. Logical models, which describe regulatory interactions qualitatively, provide a flexible and simplifying formalism to explore and understand the behaviour of a network \cite{Glass1973}. However, these approaches are unable to capture subtler features of a network that depend on specific aspects of the timing or concentration of components of the network. For this, continuous models based on, for example, ordinary differential equations (ODEs) are often employed \cite{Mogilner2006,Craciun2006}. These models describe gene regulation in much greater detail, and there are well-developed mathematical theories that provide powerful tools to distill the dynamical details of such systems. These have been used successfully to gain insight into the operation of GRNs and suggest explanations for otherwise difficult to understand behaviours, including emergent phenomena such as self-organisation, oscillations, spatial patterning, and scaling of pattern size \cite{Kondo2010,Novak2008,Perkins2006a,Umulis2010a}. A disadvantage of this approach, however, is that ODE models often require a large number of parameters, including binding affinities, degradation rates, production rates, etc. While it is often possible to numerically analyse large systems, the power of analytical tools is in many cases lost due to lack of knowledge of parameters or the complexity of the network.

Various techniques have been developed to reduce the complexity of models yet preserve specific features of the behaviour of the system. 
For Boolean networks it is possible to remove specific components while restructuring other parts of the network to conserve the logical structure, but this requires quite specific network topologies \cite{Naldi2011}.
In the case of mass action reactions, one approach to simplification is to generate intermediate species in a systematic way such as to make different systems comparable. This then allows the identification of motifs that can be simplified. This approach comes with the natural limitation that the species involved cannot be chosen \emph{a priori} and must conform to a particular network structure \cite{Feliu2013}. Other methods to reduce complexity involve removing nodes or perturbing the system and analysing \textit{a posteriori} the effect that this has on the dynamics of the network. This can be helpful to develop intuition, but generally requires the investigation of a large number of combinations of perturbations \cite{Okino1998,Gay2010}. A further approach, although applicable only in rather specific cases, is to enforce a timescale separation between network nodes that is chosen not to perturb the dynamics significantly \cite{Sunnaker2011}. One can find ``morphisms'' that relate network structure to function in order to then simplify a set of nodes into potentially simpler motifs; this can reveal an interpretation of how certain motifs work but does not provide a direct conversion from the original network \cite{Cardelli2014}. Less formal methods for replacing well understood network motifs with simpler motifs producing similar behaviour have also been extensively explored \cite{Alon2007}. 
Along similar lines, entire parts of a network can be substituted with effective nodes with more complicated dynamics that produce similar output for a set of chosen species, although the feasiblity of this depends on network structure and the complexity of the original dynamics \cite{Apri2012}. 

We focus in this paper on reducing model complexity by tracking a subnetwork that is embedded within the remainder -- or ``bulk'' -- of a larger system. Analysing the behaviour of the subnetwork and its interaction with the bulk can help reveal and rationalise the properties of the entire network. With any coarse graining approach, including those discussed above, a compromise must be reached between the precision of the method, i.e.\ how well it captures the dynamics of the full network, and the simplicity and interpretability of the resulting description. For example, the introduction of intermediate species can produce model reductions with high precision but these may not be interpretable \cite{Feliu2013}, while morphism approaches provide interpretable insights into the structure but capture the full model dynamics only for specific initial conditions and system parameters \cite{Cardelli2014}. We consider important for interpretability the ability to \emph{choose} a subnetwork, guided by biological relevance or the availability of data; none of the methods described above allow this and instead identify a subnetwork based on their internal criteria.

\begin{figure}
\centering
\includegraphics[width=1\textwidth]{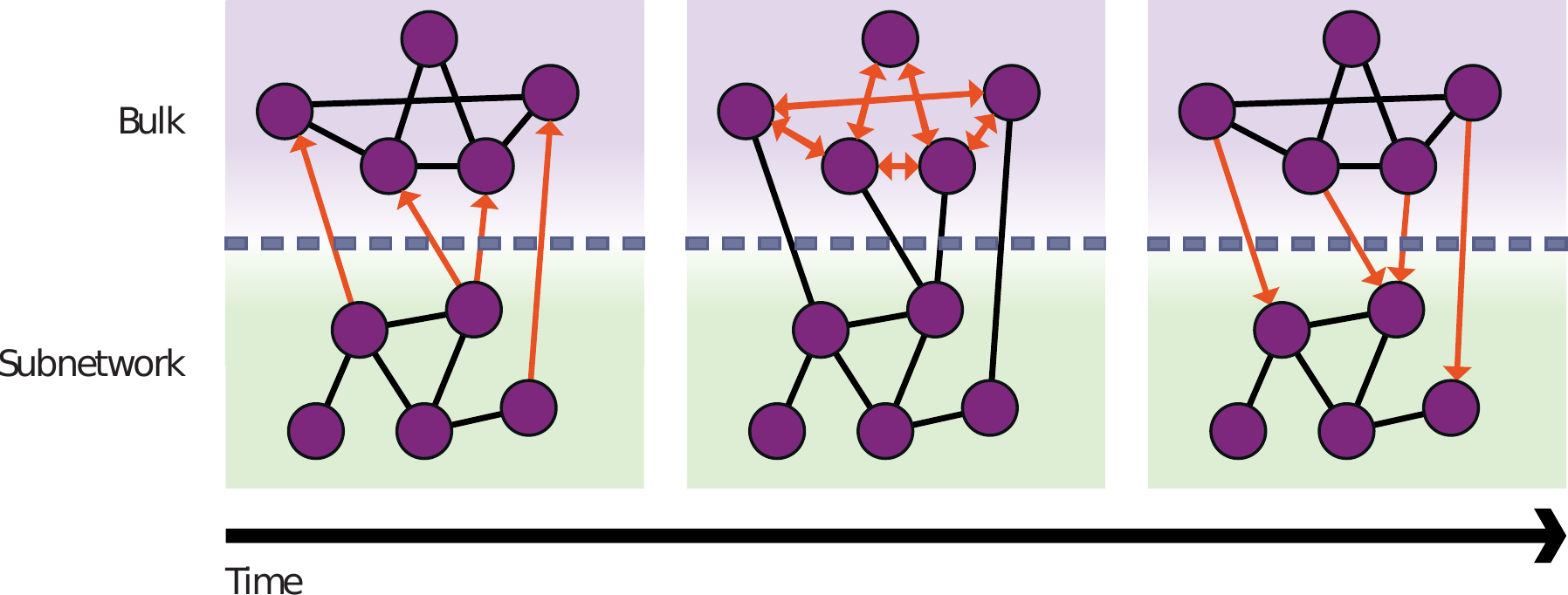}
\caption{{\bf Sketch describing the concept of memory within a reaction network.} A network is divided into bulk and subnetwork, $x$ axis represents time. Concentration changes in the subnetwork act as signals that leave the subnetwork and travel into the bulk. There they interact with other bulk species and return at a later timepoint via bulk-to-subnetwork interactions. The net effect of such interactions is thus that the subnetwork reacts to its own past. The precise influence of past subnetwork states is governed by \emph{memory functions} that depend on the time difference, \emph{i.e.}\ on how long ago the relevant signal has left the subnetwork.\label{fig:sketch}}
\end{figure}

One class of methods that offers the potential to balance accuracy and interpretability are Zwanzig-Mori projections \cite{Mori1965b,Zwanzig1961a}. Originally developed to allow the extraction of macroscopic equations from a microscopic description of the dynamics -- a brief overview can be found in \emph{e.g.}~\cite{Ritort2003} -- these methods have since been used to separate a network into an arbitarily chosen subnetwork and bulk in ways that preserve substantial features of the original temporal dynamics \cite{Rubin2014,Rubin2016}. Used in this way, the approach describes the concentration of components in the subnetwork in detail, while the activities of the species in the bulk are replaced with so-called `memory functions'. These memory functions are derived from the detailed kinetic description of the remainder of the network and summarise how, by acting via the bulk, the past behaviour of the subnetwork influences its current state (Fig.~\ref{fig:sketch}). The resulting memory functions are functions of time difference, describing the amplitude of a signal returning from the bulk some specified time after the original signal left the subnetwork.
Additionally, it is possible to separate each memory function in order to analyse through which bulk species the signal is flowing, thus providing a means to gain a more systematic understanding of a system. 

Evolving from its original applications to critical dynamics and supercooled liquids near the glass transition \cite{Gotze1992}, this approach has been applied to biochemical networks \cite{Rubin2014}, where it has been used to analyse the dynamics of signaling in the EGFR (epidermal growth factor receptor) network. It has since been developed further to include also enzymatic Michaelis-Menten reactions \cite{Rubin2016}. Here we apply the Zwanzig-Mori approach to the analysis of GRNs. We derive a general method that is applicable to any thermodynamic state ensemble representation of protein dynamics including transcriptional networks. These thermodynamic equations mechanistically represent interactions between proteins via transcriptional regulation \cite{Bintu2005a} and can be expanded into mass action reactions to which the Zwanzig-Mori projection formalism can be applied. This approach not only justifies a heuristic method where the thermodynamic equations are expanded to quadratic order around a steady state to obtain a set of effective unary and binary reactions, but also opens up the possibility of developing more advanced projection approximations. We test the method on a cross repressive motif, where we show that we are able to qualitatively reproduce the dynamics of the system and obtain a simple intuitive interpretation of the memory functions. Finally, we use the approach on a GRN from the vertebrate neural tube, a well characterised developmental GRN composed of four interacting transcription factors \cite{Cohen2014a}. This reveals the importance of specific links within the network and identifies features that appear to be present primarily to increase the robustness of the network to initial conditions, rather than maintain the steady states. Taken together, the study provides evidence that Zwanzig-Mori projections are a powerful and efficient method to simplify and explore the behaviour of gene regulatory networks, with memory functions providing new tools for probing the dynamics near steady state.

\section{Methods}

\subsection{Thermodynamic state ensemble models for gene regulation dynamics}

Several classes of model have been developed to describe gene regulation. One such class consists of thermodynamic state ensemble models in which equations are constructed that relate interactions between key components of the regulatory mechanism to gene expression \cite{Shea1985,Bintu2005a,Sherman2012}. In these models, all possible states of a regulated gene are enumerated. Each state consists of a specific set of transcription factors bound to a DNA cis-regulatory element, weighted by the affinity of interactions. The rate at which a protein is produced is then represented as a ratio of the weight of the subset of states that promote gene expression, to the total weight of all states. We define the vector $\bm{x}$ such that it contains the protein concentrations for each gene. We can then generically define the time evolution for any protein concentration, indexed by $j$, as
\begin{align}
	\partial_t x_j&=\sum_{\bm{n}} \xdna\alpha_{(j,\bm{n})}-x_j\beta_{j}\label{eq:oriP}\\
	\xdna&=\frac{w_{(j,n)}\prod_i x_i^{n_i}}{\sum_{n'}w_{(j,n')}\prod_i x_i^{n'_i}}\label{eq:DNAw}
\end{align}
Here $\beta_j$ is the decay rate of protein $j$. We use the summation over $\bm{n}$ to represent all possible binding conformations for the DNA that produces protein $j$. The term $\alpha_{(j,\bm{n})}$ represents the production rate of protein $j$ from a particular DNA conformation $\bm{n}$. The factor $w_{(j,\bm{n})}$ is the specific affinity for individual proteins to the DNA that produces species $x_j$. Finally $\xdna$ is the concentration of DNA producing protein $j$ and in conformation $\bm{n}$. This is assigned in (\ref{eq:DNAw}) as the weight of the conformation, dependent on the protein concentrations in a mass-action form, normalised by the total weight of all conformations. This ensures that $\xdna$ lies in the range $[0,1]$, so it is not an absolute concentration; any overall DNA concentration scale is therefore to be understood as incorporated into the protein production rates $\alpha_{(j,\bm{n})}$. Accordingly one can also think of $\xdna$ as the probability of finding a certain conformation $\bm{n}$ of the DNA producing protein $j$. As made explicit in (\ref{eq:DNAw}), the DNA conformation label $\bm{n}$ is a collection of integers $n_i$ counting how many copies of protein $i$ are bound to the DNA.

\subsection{Setup of projection method}

The Zwanzig-Mori method in general starts from the choice of a set of observables for which dynamical equations are to be obtained. Initially we select, as the simplest possible observables, the deviations from steady state of the concentrations of the chosen subnetwork species. For these observables the Zwanzig-Mori projection leads to a set of dynamical equations of the form
\begin{align}
	\partial_t\delta x_i =&
\sum_{j=1}^{N^s} \delta x_j\Omega_{ji}
+\int_0^t dt' \left(\sum_{j=1}^{N^s} \delta x_j(t') \underbrace{M_{ji}(t-t')}_{\text{Memory function}}
\right)
+r_i(t)\label{eq:lin_proj}
\end{align}
Here $\delta x_i$ are the subnetwork concentration deviations as defined explicitly below and $N^{\rm s}$ indicates the number of subnetwork species. 
The $\Omega_{ji}$ define a \emph{rate matrix} $\bm{\Omega}$ that represents subnetwork interactions. 
The $r_i(t)$ are so-called \emph{random forces} that arise from the uncertainty about the initial (time $t=0$) state of the bulk concentrations. (For the linear dynamics they can be expressed in closed form \cite{Rubin2014} but this is not useful in our context as discussed in Sec.~\ref{sec:NL} below.)

The key quantities in the equations (\ref{eq:lin_proj}) are the \emph{memory functions} $M_{ji}(\Delta t)$, which describe how past concentration fluctuations $\delta x_j(t')$ influence the current time evolution. However, while the projected equations are formally exact, it is in general impossible to evaluate these memory functions explicitly. One scenario where this \emph{can} be done is a network of unary reactions, where the full reaction equations are linear in the concentrations \cite{Rubin2014}. This suggests a heuristic approach to calculating memory functions in the current context, which is to linearize the thermodynamic equations. One can think of this as generating a set of effective unary reactions (at least loosely; see Sec.~\ref{sec:justification}). Explicitly, one linearly expands the time evolution equations (\ref{eq:oriP}) with (\ref{eq:DNAw}) inserted. The expansion is performed around a steady state with concentrations $y_i$, \emph{i.e.}\ in terms of the deviations $\delta x_i=x_i-y_i$. The r.h.s.\ of (\ref{eq:oriP}) then contains only linear terms in the $\delta x_i$ since constant terms cancel out because of the steady state condition. One can therefore write the time evolution equations in matrix form~\cite{Rubin2014}
\begin{equation}
\partial_t \delta x_i = \sum_{j=1}^{N^{\rm s}}\delta x_j L_{ji}
\end{equation}
for an appropriately defined matrix $\LM{}$ with entries $L_{ji}$.

\subsection{Projected equations for linearised dynamics \label{sec:linear}}

From the matrix $\LMb{}$ one can calculate the terms in the (linear) projected dynamics of the subnetwork concentrations from (\ref{eq:lin_proj}) as shown in~\cite{Rubin2014}. We summarize the method here.
One separates the set of all (linear in concentration) observables into subnetwork (s) and bulk (b), and assumes as above that these concentration deviations $\delta x_i$ are numbered so that the first $N^{\rm s}$ are the subnetwork observables while the rest are the bulk observables. The matrix $\LM{}$ for the linearised dynamics then separates into blocks according to the different observables:
\begin{align}
	\LMb{}&=
	\begin{pmatrix}	
	\LMb{S,S}&\vline&\LMb{S,B}\\
	\hline\rule{0pt}{3ex}
	\LMb{B,S}&\vline&\LMb{B,B}
	\end{pmatrix}\label{eq:matL_PD}
\end{align}
where labels S and B represent respectively the set of species in the subnetwork and in the bulk.

We use the labels $\textrm{S}$ and $\textrm{B}$ here to represent subnetwork and bulk species respectively, this will be useful once more subnetwork and bulk observables arise (see Sec.~\ref{sec:tec}).
The rate matrix $\bm{\Omega}$ with entries $\Omega_{ji}$ in the projected equations (\ref{eq:lin_proj}) is then simply the subnetwork block of $\LMb{}$,
\begin{equation}
\bm{\Omega}=\LMb{S,S}\label{eq:omega}
\end{equation}
The memory functions similarly form the elements of a memory matrix, which can be expressed in terms of the blocks of $\LMb{}$ as \cite{Rubin2014}
\begin{align}
	\bm{M}(\Delta t)=\LMb{S,B}e^{\LMb{B,B}\Delta t}\LMb{B,S}\label{eq:ZM_form}
\end{align}
This form allows straightforward evaluation of the memory functions starting from any given matrix $\LMb{}$ describing the linearised dynamics.

\subsection{Nonlinear projected equations\label{sec:NL}}

To capture the leading nonlinear corrections to the subnetwork dynamics, it is natural to enlarge the set of subnetwork observables by adding quadratic observables, i.e.\ products of concentration deviations. The projected equations then become
\begin{align}
	\partial_t\delta x_i =&
\sum_{j=1}^{N^s} \delta x_j\Omega_{ji}
+\sum_{1\leq j\leq k\leq N^s}^{N^s} \delta x_j\delta x_k\Omega_{(jk)i}\\
&+\int_0^t dt' \left(\sum_{j=1}^{N^s} \delta x_j(t')\underbrace{{M_{ji}(t-t')}}_{\text{Linear memory}}
+\sum_{1\leq j\leq k\leq N^s}^{N^s} \delta x_j(t')\delta x_k(t')\underbrace{{M_{(jk)i}(t-t')}}_{\text{Nonlinear memory}}\right)\nonumber\\
&+r_i(t)\nonumber
\end{align}
This form is analogous to (\ref{eq:lin_proj}) but now includes the quadratic observables that we have retained, both in the subnetwork interactions (rate matrix terms) and in the memory terms. Once again $r_i(t)$ represents the \emph{random forces} but in this nonlinear case further characterization would require knowledge of the statistics of the initial bulk fluctuations, which is not generally available. We therefore disregard these terms, making (\ref{eq:lin_proj}) a \emph{closed} system of equations for the time evolution of the subnetwork concentrations. Note that neglecting the random forces is equivalent to assuming that the bulk concentrations are at their steady state values at the initial time. Formally these initial bulk concentrations should be random rather than deterministic to make the Zwanzig-Mori projection well defined; we follow the strategy of~\cite{Rubin2014} here and use initial Poisson distributions in the limit of vanishing variance. This limit has already been taken in the results above~\cite{Rubin2014}. How large protein copy numbers have to be to allow stochastic effects to be safely neglected is something that could be studied in future work; the answer will depend, among other things, on the degree of nonlinearity of the GRN time evolution equations~\cite{Thomas2012c}.

Moving on to the nonlinear memory functions, we are again faced with the difficulty that these cannot in general be evaluated explicitly. However, previous work~\cite{Rubin2014} has developed a systematic approximation technique, which is applicable to the case of reaction networks with unary and binary reactions described by mass action kinetics. 
In such networks the time evolution equations for linear observables have linear and quadratic terms on the r.h.s.\ Inserting these into the equations for \emph{quadratic} observables using the product rule gives \emph{e.g.}
\begin{align}
	\partial_t(\delta x_1 \delta x_2) = (\partial_t x_1) \delta x_2 + \delta x_1 (\partial_t\delta x_2)
\end{align}
and one sees that cubic terms arise. The approximation in~\cite{Rubin2014} neglects these terms in the spirit of an expansion to second order in the changes from steady state $\delta x$. These changes are therefore implicitly assumed to be small.
Defining a vector $\bm{z}$ that collects all $\delta x$ variables and their products such as $\delta x_1^2$, $\delta x_1 \delta x_2$, one can now again write a matrix form of the time evolution equations:
\begin{equation}
\partial_t z_\alpha = \sum_\beta z_\beta L_{\beta\alpha}
\label{eq:Ldef}
\end{equation}
The rate matrix entries and memory functions can then be calculated~\cite{Rubin2014} from the general formulae (\ref{eq:omega},\ref{eq:ZM_form}), applied to the expanded $\bm{L}$ matrix constructed as explained above. Because the subnetwork block S now contains linear \{s\} and quadratic observables \{ss\}, the matrices have the corresponding block structures. The coefficients $\Omega_{ji}$ are collected in the block $\bm{\Omega}^{\rm s,s}$, for example, and the $\Omega_{(jk)i}$ in the block $\bm{\Omega}^{\rm ss,s}$ of the overall rate matrix $\bm{\Omega}$. The memory functions $M_{ji}(\Delta t)$ and $M_{(jk)i}(\Delta t)$ are similarly contained in blocks $\bm{M}^{\rm s,s}$ and $\bm{M}^{\rm ss,s}$ of the memory matrix $\bm{M}$.

To apply the above technique in order to derive nonlinear projected subnetwork equations for GRN dynamics, the obvious heuristic route is again to expand the dynamical equations (\ref{eq:oriP}), this time to second order in the $\delta x_i$, to obtain an effective set of unary and binary reactions. This set defines the matrix $\LMb{}$, from which the rate matrix and memory functions can then be found as explained above.

\subsection{Justifying the heuristics: network expansion}
\label{sec:justification}

The heuristic method set out above for deriving linear or nonlinear subnetwork equations for GRN dynamics is a useful practical recipe, but its mathematical basis is not obvious. One difficulty is that the thermodynamic GRN equations (\ref{eq:oriP}) involve rational functions of the concentrations such as $1/(1+x_1)$, and while these can be Taylor expanded they have only finite radius of convergence. Not only does the heuristic method then have to throw away an infinite number of higher order terms when truncating the Taylor expansion after linear or quadratic order, but it may implicitly be using the resulting approximation outside the regime where the expansion is even convergent. In addition, while we have loosely talked about the expanded time evolution equations as representing networks of effective reactions, such an interpretation is in no way assured as there is no underlying mechanistic model and hence no constraints that would ensure an appropriate stoichiometry or even positivity of rate constants.

Our mathematical contribution in this paper is to demonstrate that the above limitations can be resolved, and the heuristic approach justified, by initially \emph{expanding}
the thermodynamic equations (\ref{eq:oriP},\ref{eq:DNAw}) into a network of well-defined unary and binary reactions. The way to achieve this is already suggested by (\ref{eq:DNAw}): we represent each possible existing DNA conformation as an additional species whose concentration has its own dynamical evolution, such that in an appropriate limit of fast DNA binding and unbinding, it reaches the quasi-steady state (QSS) values of the DNA conformation concentrations (\ref{eq:DNAw}). The prefix ``quasi'' refers to the fact that this steady state is for given protein concentrations, which themselves change slowly in time. We can then apply the Zwanzig-Mori method to this network to extract memory functions for the chosen subnetwork. In doing this we can justify removing higher order terms as in~\cite{Rubin2014} as the system is composed of only first and second order reactions. Finally the limit of fast DNA binding and unbinding has to be taken. Our result is that this procedure results in exactly the same projected equations as the heuristic approach described above, thus putting the method on a firm mathematical footing. The conclusion applies not only in the context of GRNs but in all networks with the appropriate timescale separation, provided that all fast observables are treated as part of the bulk. More sophisticated projection methods can then be derived by retaining some fast observables (in our case, subnetwork DNA) in the subnetwork. For the case of Michaelis-Menten dynamics, previous work has shown that such an approach allows one to derive projected equations that retain all subnetwork nonlinearities~\cite{Rubin2016}. 
More generally, by embedding GRN models based on thermodynamic state ensembles into the broad class of mass action reaction networks, our network expansion method opens up the possibility of incorporating stochastic effects and of applying a broad range of other approximation and model reduction techniques~\cite{Schnoerr2017}.

\section{Results}

In this section we first provide our mathematical results, where we use a network expansion technique to provide a justification for the heuristic approach to finding memory functions for GRNs that was described in Secs.~\ref{sec:linear} and~\ref{sec:NL} above. We then demonstrate applications of the approach with case studies. We first illustrate the memory function method on a simple example system. Finally, the results from the biological application to neural tube patterning are given in Sec.~\ref{sec:app} onwards.

\subsection{Mathematical results\label{sec:tec}}

\subsubsection{Network expansion applied to a protein-DNA binding mechanism}

\begin{figure}
\centering
\includegraphics[width=1\textwidth, height=1\textwidth, keepaspectratio]{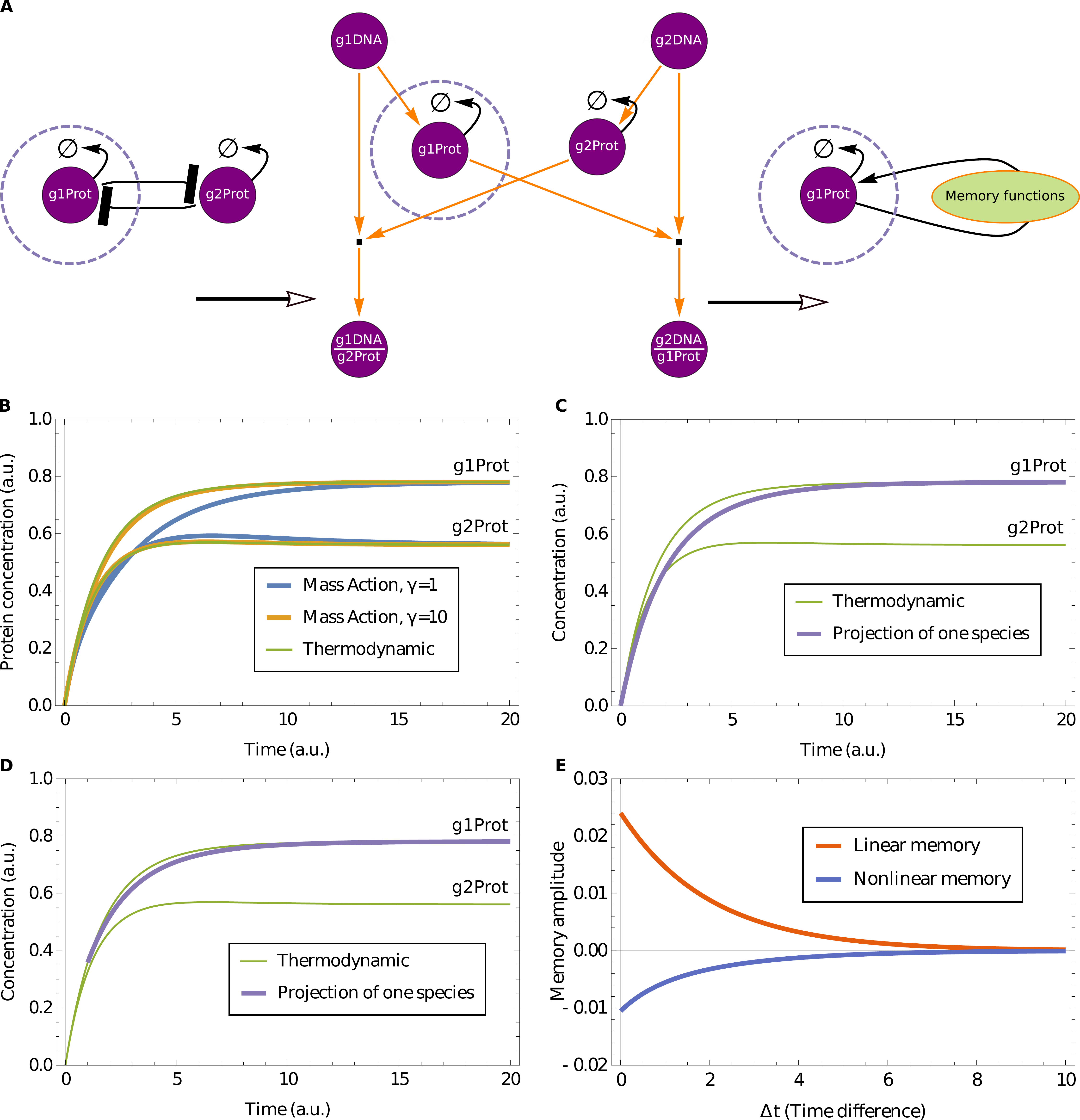}
\caption{{\bf Example of application of Zwanzig-Mori projection.} (\textbf{A}) Illustration of the methodology, for the example of a cross repressive motif.  First the nonlinear thermodynamic reactions (left) are expanded into mass action reactions with an appropriate timescale separation (centre). This generates additional nodes that represent the possible DNA conformations for both proteins, with \emph{e.g.}\ g1DNA/g2Prot indicating DNA for gene 1 with protein 2 bound to it.
To the expanded network we can apply the projection approach, retaining only the concentration of protein 1 in the subnetwork. The effect of the rest of the network -- the bulk -- is captured via memory terms (right). 
(\textbf{B}) Comparison of the cross repressive motif described via the original thermodynamic equations and the expanded mass action equations with and without timescale separation. Already for a moderate fast rate factor of $\gamma=10$ the mass action and thermodynamic time evolutions are visually indistinguishable. Two time courses are plotted in Fig.~(\textbf{B}-\textbf{D}), for g1Prot \& g2Prot. (\textbf{C}-\textbf{D}) Demonstration of the projection approach with g1Prot in subnetwork and g2Prot in bulk. The projected equations track the dynamics of the original thermodynamic equation with a reduced system that contains memory functions. However, one can observe in (\textbf{C}) that accuracy can be lost in the transient if the system is initiated too far away from the fixed steady state. (\textbf{D}) Starting nearer to the steady state (at $t=1$) substantially increases the accuracy of the projected description. (\textbf{E}) Example of memory functions of protein 1 to itself, decaying with time difference. The linear memory function is positive as expected from the network (positive feedback loop), while the nonlinear term is negative to correct for range-limiting nonlinearities that the linear terms cannot capture. Parameters used are $\alpha_1=1$, $\alpha_2=1$, $w_{\text{p1}}=1$, $w_{\text{p2}}=2$, $w_1=2$, $w_2=2$, $\beta_\text{1}=1/2$, $\beta_\text{2}=1/2$.} \label{fig:Expansion}
\end{figure}

Our network expansion approach necessitates defining new reaction rates for binding and unbinding of protein to/from DNA that are consistent with the affinities $w$ from the thermodynamic description. To be explicit, we define $k_{(j,\bm{n})}^{p+}$ as the rate constant for protein $p$ to bind to DNA coding for protein $j$ and in binding conformation $\bm{n}$, with a similar meaning for $k_{(j,\bm{n})}^{p-}$. These rates will thus describe the interaction each DNA species has with the protein species that bind to it.
With these rate constants we write down an equation for the time evolution of each DNA concentration $\xdna$:
\begin{align}
	\partial_t\xdna&=\sum_p\left(
	k_{(j,\bm{n}-\ep)}^{p+}x_{(j,\bm{n}-\ep)}x_p
	-k_{(j,\bm{n})}^{p+}\xdna x_p
	+k_{(j,\bm{n}+\ep)}^{p-}x_{(j,\bm{n}+\ep)}
	-k_{(j,\bm{n})}^{p-}x_{(j,\bm{n})}
	\right)\label{eq:meDNA}
\end{align}
The first term on the r.h.s.\ of (\ref{eq:meDNA}) represents binding of protein $p$ to DNA conformation $(j,\bm{n}-\ep)$ to make DNA conformation $(j,\bm{n})$. Here $\ep$ is a unit vector whose $i$-th entry is $\delta_{ip}$, \emph{i.e.}\ a 1 at position $p$ and zeros elsewhere. Similarly the third term on the right of (\ref{eq:meDNA}) describes unbinding of protein $p$. The other two terms capture how DNA in conformation $(j,\bm{n})$ can be lost by these two processes.

We next establish the relationship between the rate constants $k$ in the mass action representation (\ref{eq:meDNA}) and the affinities $w$ of the thermodynamic form. The rate constants need to be chosen to ensure the QSS (\ref{eq:DNAw}) of the DNA concentrations. The simplest assumption is detailed balance, which requires that in this steady state the net change in DNA concentration from any \emph{individual pair} of binding and unbinding reactions between two conformations vanishes. Detailed balance means, in particular, that where there are multiple reaction paths from one DNA conformation to another, which might be distinguished by the order in which the various proteins bind, each path carries zero reaction flux rather than having a preferred forward or backward reaction direction.  Balancing in this way the net change from binding of protein $p$ to DNA in conformation $(j,\bm{n})$ with the reverse unbinding transition, for the QSS concentrations (\ref{eq:DNAw}), gives
\begin{align}
k_{(j,\bm{n})}^{p+} x_p w_{(j,\bm{n})} \prod_i x_i^{n_i} 
= k_{(j,\bm{n}+\ep)}^{p-} w_{(j,\bm{n}+\ep)} \prod_i x_i^{n_i+\delta_{ip}} 
\end{align}
The protein concentrations cancel from this as they should and we are left with a constraint on the rate constants:
\begin{align}
	\frac{k_{(j,\bm{n})}^{p+}}{k_{(j,\bm{n}+\ep)}^{p-}}=
	\frac{w_{(j,\bm{n}+\ep)}}{w_{(j,\bm{n})}}\label{eq:defK}
\end{align}
To finish our construction, we need to ensure that the DNA concentrations reach the QSS values that we have imposed above with the detailed balance constraint (\ref{eq:defK}). This requires that we make all DNA binding and unbinding reactions fast. Formally we introduce a fast rate factor $\gamma$ and write all rate constants in the form $k^{p\pm}_{(j,\bm{n})}=\gamma\tilde{k}^{p\pm}_{(j,\bm{n})}$, with the understanding that we will take the limit $\gamma\to\infty$ at fixed $\tilde{k}$. This construction is analogous to the one previously used for Michaelis-Menten reactions \cite{Rubin2016}. As there, we also need to ensure that the changes to protein concentrations from binding to DNA are negligible, as such effects are not described by the original thermodynamic equations (\ref{eq:oriP}). (We do, however, include them in the finite-$\gamma$ numerics of Fig.~\ref{fig:Expansion}B.)
This is achieved by scaling all DNA concentrations as $x_{(j,\bm{n})}=\tilde{x}_{(j,\bm{n})}/\gamma'$. The intuition for this construction is that the amount of protein bound to DNA is, at most, of the order of the total concentration of DNA, which vanishes for $\gamma'\to\infty$. Finally, to compensate for the DNA concentration scaling, protein production rates have to be scaled as $\alpha_{(j,\bm{n})}=\gamma' \tilde{\alpha}_{(j,\bm{n})}$. 

With the above rescalings, the limit $\gamma\to\infty$ ensures that the ``fast'' species, \emph{i.e.}\ the DNA concentrations are always in QSS with respect to the ``slow'' protein concentrations. The limit $\gamma'\to\infty$, on the other hand, means that the effects of binding to, and unbinding from, DNA can be neglected in the equations for the protein concentrations. These statements can be shown mathematically along the lines of the arguments in \cite{Rubin2016}. After dropping all tildes on rates and concentrations again -- this is the notational convention we adopt for the rest of the paper -- the time evolution of the concentrations in our expanded network is given by (\ref{eq:oriP}) for the protein equations, while for the DNA species one has
\begin{align}
	\partial_t\xdna&=\gamma \sum_p\left(
	k_{(j,\bm{n}-\ep)}^{p+}x_{(j,\bm{n}-\ep)}x_p
	- k_{(j,\bm{n})}^{p+}\xdna x_p
	+ k_{(j,\bm{n}+\ep)}^{p-}x_{(j,\bm{n}+\ep)}
	- k_{(j,\bm{n})}^{p-}x_{(j,\bm{n})}
	\right)\label{eq:meDNAG}
\end{align}
Note that in the above argument we do not require a specific relation between $\gamma$ and $\gamma'$. In reality the rate for producing a single protein from the appropriate DNA, which is of order $\gamma'$, is lower than the DNA binding/unbinding rates, which are $O(\gamma)$. However, quantifying this would be non-trivial as the model described here simplifies many biological steps into each of the two elementary processes of protein production and binding/unbinding. In any event, for our argument to hold we only need both $\gamma$ and $\gamma'$ to be large, and for simplicity we take then $\gamma=\gamma'$ as was done in \cite{Rubin2016}.

\subsubsection{Obtaining linearised projected equations}
We now proceed to use the expanded network to derive linearised projected equations. After the expansion, this requires us to keep track of deviations from steady state of all protein and DNA concentrations. We label these observables by ``s'' and ``b'' for subnetwork and bulk proteins as before, and ``a'' for DNA. Below we will refer to the DNA species in the system as ``fast'', as expressed mathematically by the factor $\gamma$ on the r.h.s.\ of (\ref{eq:meDNAG}), while protein species are ``slow''.
Note that the \emph{rate constants} for protein production are also fast, of order $\gamma'$, but that the resulting \emph{changes} to protein concentrations are slow because of the low concentration of (protein-producing) DNA. 

We construct our $\LMb{}$ matrix by placing the DNA in the bulk, such that $\textrm{S}=\{\textrm{s}\}$ and $\textrm{B}=\{\textrm{b}, \textrm{a}\}$, giving the block form
\begin{align}
	\LMb{}&=
	\begin{pmatrix}	
	\LMb{S,S}&\vline&\LMb{S,B}\\
	\hline\rule{0pt}{3ex}
	\LMb{B,S}&\vline&\LMb{B,B}
	\end{pmatrix}=\begin{pmatrix}	
	\LMb{s,s}&\vline&\LMb{s,b}&\LMb{s,a}\\
	\hline\rule{0pt}{3ex}
	\LMb{b,s}&\vline&\LMb{b,b}&\LMb{b,a}\\
	\LMb{a,s}&\vline&\LMb{a,b}&\LMb{a,a}
	\end{pmatrix}\label{eq:matL_FS}
\end{align}
The expression (\ref{eq:ZM_form}) for the resulting memory function now explicitly involves the fast rate factor $\gamma$ and the rate constants $\tilde{k}^{p\pm}_{(j,\bm{n})}$ used in the construction of our expanded network. These parameters appear in the third column in the second expression for $\LMb{}$ in (\ref{eq:matL_FS}), which encodes the time evolution equations for the DNA species. Our remaining task is to take the limit $\gamma\to\infty$, with the aim of obtaining an expression for the memory function that only involves parameters of the original thermodynamic equations (\ref{eq:oriP},\ref{eq:DNAw}). To take this limit, one notes that all blocks in the third column of (\ref{eq:matL_FS}) are proportional to $\gamma$. As the memory function contains an exponential of $\LMb{}$, it will then have contributions decaying for time differences of order $1/\gamma$, arising from the DNA dynamics, as well as slow contributions decaying on timescales of order unity.

For $\gamma\to\infty$ the fast memory function contributions become effectively instantaneous and add to the rate matrix, and only the slow contributions remain in the memory function. This separation into fast and slow pieces of the memory can be performed using the method in \cite{Rubin2016} because $\LMb{}$ has exactly the same division into fast and slow blocks as there. It leads to the simple result that the final rate matrix and (slow) memory function for $\gamma\to\infty$ can be calculated from an effective $\LMb{}$-matrix no longer involving the fast degrees of freedom.
To state the expression for the effective $\LMb{}$, we first isolate the slow piece of our original $\LMb{}$:
\begin{align}
	\bm{L}_{\backslash a}=
	\begin{pmatrix}
	\LMb{s,s}&\vline&0\\
	\hline\rule{0pt}{3ex}
	0&\vline&\LMb{b,b}
	\end{pmatrix}\label{eq:noFMemL}
\end{align}
(The zero blocks here arise from the fact that subnetwork and bulk proteins do not interact directly in our GRN setting, but only via the DNA species.) 
The effective $\LMb{}$-matrix that gives us the rate matrix and memory function for $\gamma\to\infty$ is then of the form $\bm{L}_{\text{eff}} = \bm{L}_{\backslash a} + \Delta\bm{L}_{\backslash a}$, which written out reads:
\begin{align}
\bm{L}_{\text{eff}} = 
	\begin{pmatrix}
	\LMb{s,s}&\vline&0\\
	\hline\rule{0pt}{3ex}
	0&\vline&\LMb{b,b}
	\end{pmatrix}
	-\begin{pmatrix}
	\LMb{s,a}(\LMb{a,a})^{-1}\LMb{a,s}&\vline&\LMb{s,a}(\LMb{a,a})^{-1}\LMb{a,b}\\
	\hline\rule{0pt}{3ex}
	\LMb{b,a}(\LMb{a,a})^{-1}\LMb{a,s}&\vline&\LMb{b,a}(\LMb{a,a})^{-1}\LMb{a,b}
	\end{pmatrix}\label{eq:effL}
\end{align}
Here the second term including the minus is the additional contribution $\Delta\bm{L}_{\backslash a}$ from the fast degrees of freedom. The matrix $\bm{L}_{\text{eff}}$ can be inserted directly into (\ref{eq:omega},\ref{eq:ZM_form}) to obtain the projected equations for linearised dynamics, with the fast rate limit already taken. 

\subsubsection{Fast binding limit as Quasi-Steady State elimination}

As in \cite{Rubin2016}, the result (\ref{eq:effL}) has a simple interpretation: $\bm{L}_{\text{eff}}$ can be obtained by eliminating the fast degrees of freedom from the linearised time evolution equations using a QSS condition. Using the block form (\ref{eq:matL_PD}), these equations can be written as
\begin{subequations}
\begin{align}
	\partial_t\dx{s}&=
	\dx{s}\LMb{s,s}+
	\dx{a}\LMb{a,s}\label{eq:timeEva}\\
	\partial_t\dx{b}&=
	\dx{b}\LMb{b,b}+
	\dx{a}\LMb{a,b}\label{eq:timeEvb}\\
	\partial_t\dx{a}&=
	\dx{s}\LMb{s,a}+
	\dx{b}\LMb{b,a}+
	\dx{a}\LMb{a,a}\label{eq:timeEvc}
\end{align}
\end{subequations}
where $\dx{s}$, $\dx{b}$ and $\dx{a}$ are vectors collecting the concentration deviations of subnetwork proteins, bulk proteins and DNA species, respectively. 
All blocks of $\LMb{}$ appearing in (\ref{eq:timeEvc}) are proportional to $\gamma$. This justifies elimination of the fast variables using a QSS condition for these fast (DNA) degrees of freedom, effectively setting the r.h.s.\ of (\ref{eq:timeEvc}) to zero. The fast species concentrations are then expressed as:
\begin{align}
	\dx{a}=
	-\left(\dx{s}\LMb{s,a}+
	\dx{b}\LMb{b,a}\right)
	(\LMb{a,a})^{-1}\label{eq:DNAQSS}
\end{align}
If we now substitute this back into the time evolution equations (\ref{eq:timeEva},\ref{eq:timeEvb}) we obtain:
\begin{align}
	\partial_t\dx{s}&=
	\dx{s}(\LMb{s,s}-\underbrace{\LMb{s,a}(\LMb{a,a})^{-1}\LMb{a,s}}_{\Delta\LMb{s,s}})+
	\dx{b}(\underbrace{-\LMb{b,a}(\LMb{a,a})^{-1}\LMb{a,s}}_{\Delta\LMb{b,s}})\label{genSlow}
	\\
	\partial_t\dx{b}&=
	\dx{s}(-\LMb{s,a}(\LMb{a,a})^{-1}\LMb{a,b})+
	\dx{b}(\LMb{b,b}-\LMb{b,a}(\LMb{a,a})^{-1}\LMb{a,b})\nonumber
\end{align}
Writing these time evolution equations in matrix form gives exactly the effective $\LMb{}$-matrix $\bm{L}_{\text{eff}}$ defined in (\ref{eq:effL}), as claimed. Two terms have been highlighted here by brackets for later comparison.

\subsubsection{Equivalence to heuristic linearisation \label{sec:equivLin}}

So far we have shown that the rate matrix and memory functions for the linearised dynamics can be found by applying standard projection results to a set of time evolution equations for only the slow degrees of freedom. This reduced set of equations is obtained by first expanding the thermodynamic equations into a set of mass-action equations, linearising, and finally eliminating the fast variables using a QSS assumption. We next show that this procedure is equivalent to directly linearising the original thermodynamic equations, for any case where there is a timescale separation. This generalises previous work~\cite{Rubin2016} that only considered separate (non-interacting) fast species. To verify the equivalence, it is useful to write the expanded mass-action equations in the generic form
\begin{align}
	\partial_t x_s=R_s(\x{s},\x{b},\x{a}),\enspace
\enspace	\partial_t x_b=R_b(\x{s},\x{b},\x{a}),\enspace\enspace
	\gamma^{-1} \partial_t x_a=R_a(\x{s},\x{b},\x{a})\label{eq:TD_QSS}
\end{align}
where $x_s$, $x_b$ and $x_a$ are generic components of $\x{s}$, $\x{b}$ and $\x{a}$, respectively. This expanded description is constructed so that for $\gamma\to\infty$, when the $x_a$ can be replaced by their QSS values $x_a^*$, one recovers the original thermodynamic equations. These can therefore be written as
\begin{align}
	\partial_t x_s=R_s(\x{s},\x{b},\x{a}{}^*)\\
	\partial_t x_b=R_b(\x{s},\x{b},\x{a}{}^*)
\end{align}
where the dependence of the $x_a^*$ on $\x{s}$ and $\x{b}$ is defined implicitly by
\begin{equation}
R_a(\x{s},\x{b},\x{a}{}^*) = 0
\label{qss_condition}
\end{equation}
Expanding now the thermodynamic equations to linear order around a steady state one has
\begin{align}
	\partial_t\delta x_s =&
	\sum_{s'}\frac{\partial R_s}{\partial x_{s'}}
\delta x_{s'}
	+\sum_{b}\frac{\partial R_s}{\partial x_{b}}
\delta x_{b}\label{linearised_thermodynamic}\\
&        +\sum_{s'}\underbrace{\sum_{a}\frac{\partial x_a^*}{\partial x_{s'}}\frac{\partial R_s}{\partial x_a}}_{\Delta\LMb{s,s}}\delta x_{s'}
	+\sum_{b}\underbrace{\sum_{a}\frac{\partial x_a^*}{\partial x_{b}}\frac{\partial R_s}{\partial x_a}}_{\Delta\LMb{b,s}}\delta x_{b}\nonumber
\end{align}
Here all derivatives are evaluated at the steady state, and the terms in the second line arise from the variation of the QSS values of the fast variables. The required coefficients of the type $\frac{\partial x_a^*}{\partial x_{s}}$ can be found by differentiating (\ref{qss_condition}) with respect to the relevant slow variable, here $x_s$:
\begin{equation}
0 = \frac{\partial R_a}{\partial x_s} + \sum_{a'} \frac{\partial R_a}{\partial x_{a'}}\frac{\partial x_{a'}^*}{\partial x_s}
\label{eq:linearised_qss}
\end{equation}
We can now show that (\ref{linearised_thermodynamic}) is identical to (\ref{genSlow}) derived above. To see this, we note that by linearising the expanded mass action equations (\ref{eq:TD_QSS}) one can identify the entries of $\LMb{}$: for example, $\LMb{b,s}$ has entries $\partial R_s/\partial x_b$, $\LMb{s,a}$ has entries $\gamma\, \partial R_a/\partial x_s$ etc. The QSS coefficients arising from (\ref{eq:linearised_qss}) can then be written in matrix form as
$\partial x_a^*/\partial x_{s}=-(\LMb{s,a}(\LMb{a,a})^{-1})_{sa}$,  with a similar expression for $\partial x_a^*/\partial x_{b}$. Inserting these expressions into (\ref{linearised_thermodynamic}) does indeed lead directly to (\ref{genSlow}), with matching terms indicated by the brackets. An exactly analogous argument demonstrates the equivalence for $\partial_t \delta x_b$. 

With the above arguments we have justified the heuristic method of Sec.~\ref{sec:linear} for obtaining linearised projected equations for the dynamics of a subnetwork within a larger GRN written in thermodynamic form: expand the GRN equations to linear order in protein concentrations around a steady state. Then construct from this the matrix $\bm{L}_{\text{eff}}$ and partition this into ``s'' and ``b'' blocks according to the chosen subnetwork--bulk split. Finally determine the rate matrix and memory function matrix from (\ref{eq:omega},\ref{eq:ZM_form}). We invoked an expanded network of binary reactions involving DNA conformations to \emph{derive} that this is the correct method, but note that its \emph{application} only requires the original thermodynamic equations as input. Therefore the resulting projected equations are, as they should be, independent of the details of the rates $\tilde{k}^{p\pm}_{(j,\bm{n})}$ used in the construction of the expanded network. 

\subsubsection{Derivation of nonlinear projected equations\label{sec:nl}}

To capture more of the nonlinearity inherent in GRNs one can extend
the approach described in the previous sections, to include terms that are quadratic in the deviations from a steady state. The memory function then contains fast and slow contributions, with the former being transformed into additional rate matrix terms when $\gamma\to\infty$. The resulting rate matrix and slow memory functions can be obtained from an effective $\LMb{}$-matrix involving only slow (linear and quadratic) observables. As before, this $\bm{L}_{\text{eff}}$ can be viewed as arising from a QSS elimination of the fast observables. Finally, one can show that this explicit elimination of the fast observables is equivalent to directly expanding the original thermodynamic equations to \emph{quadratic} order in concentration deviations from the steady state: this justifies the heuristic method given in the Sec.~\ref{sec:NL}.

Except for the last step, the above chain of argument is analogous to the linearised dynamics case, being based entirely on the identical structure of the partition -- compare (\ref{eq:matL_PD}) and (\ref{app:A}, Eq. S.1) -- of $\LMb{}$ into fast and slow blocks. The final reduction to a quadratic expansion of the thermodynamic equations is more subtle because the projection approach treats quadratic observables as distinct and not directly tied to products of linear observables. We defer the details to \ref{app:A} and note only that our derivation there both simplifies and generalises that in \cite{Rubin2016}, from enzyme species that do not interact with each other to arbitrary interactions of the fast degrees of freedom. 

\begin{figure}
\centering
\includegraphics[width=1\textwidth, height=0.8\textwidth, keepaspectratio]{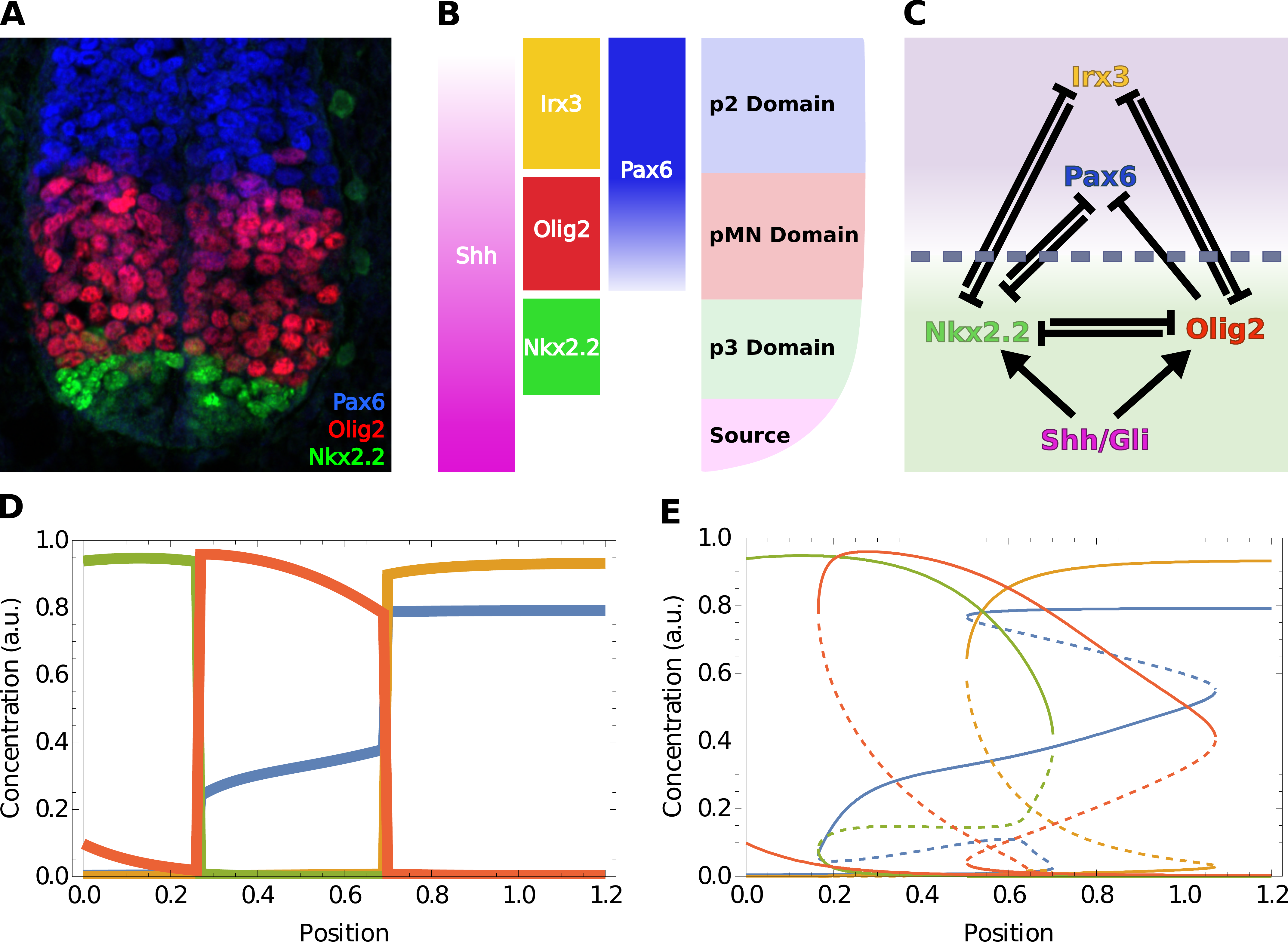}
\caption{{\bf Patterning of the vertebrate neural tube.} (\textbf{A}) Antibody staining of Wild type (WT) mouse stained for three of the main bands in dorso-ventral patterning. (Image provided by Katherine Exelby). (\textbf{B}) Illustration of neural tube patterning: ventral Shh secreted from the notocord and floor plate (termed "Source") generates patterned domains along the dorso-ventral axis. Each domain is defined by the expression of a characteristic set of genes. (\textbf{C}) GRN that patterns the three most ventral domains of the neural tube. The chosen separation of bulk (purple) and subnetwork (green) in the application of the Zwanzig-Mori projection is also shown. (\textbf{D}) Simulations of steady state pattern along the dorsoventral axis using a set of thermodynamic equations of the form of equations (\ref{eq:oriP},\ref{eq:DNAw}). These equations were taken -- along with appropriate initial conditions of 0 for all species -- from \cite{Cohen2014a}. For all plots where $x$ axis represents neural tube position, zero corresponds to the most ventral point. (\textbf{E}) Full bifurcation diagram illustrating the multistable nature of the network. Shown are steady state concentrations of the four molecular species against neural tube position, with unstable steady states marked dashed. Colours in (\textbf{D}, \textbf{E}) identify genes/proteins in the same way as in the labelling of the illustration (\textbf{B}) and of the network nodes in (\textbf{C}). This colour code is used throughout the paper unless otherwise noted.
}
\label{fig:PONI_SSs}
\end{figure}

\subsection{Case studies\label{sec:cs}}
\subsubsection{Zwangzig-Mori projection of a cross-repression motif\label{sec:ex}}

We applied the Zwanzig-Mori projection method to a cross repressive motif to illustrate our approach. Fig.~\ref{fig:Expansion}A shows the expansion procedure for this very simple regulation network: each of two genes is repressed through one binding site for the other factor. This system can be expanded to represent all possible DNA conformations, bound and unbound, as shown graphically by the four outer nodes in the network in the centre of Fig.~\ref{fig:Expansion}A. As the arrows indicate, in the expanded network unbound DNA can bind protein to form bound DNA; the reverse process is also possible. In this way one creates a set of reaction equations that has been extended and brought into mass action form by including the concentration of DNA conformations. 
The equations used for Fig.~\ref{fig:Expansion} are of the form of (\ref{eq:oriP}) \& (\ref{eq:DNAw}):
\begin{align}
\partial_t[\text{g1Prot}]={\alpha_1}\left({w_1+w_{\text{p2}}[\text{g2Prot}]}\right)^{-1}-\beta_\text{1}[\text{g1Prot}]
\end{align}
and conversely for g2Prot.
In numerical simulations of the expanded network we first checked the fast rate limit, which we control by setting the ratio, $\gamma$, of typical DNA reaction rates to protein production rates. Fig.~\ref{fig:Expansion}B shows that already with a moderate value of $\gamma=10$ the predicted time courses of protein concentrations are almost identical to those predicted by the thermodynamic equations. As $\gamma\to\infty$ the trajectories from the mass action equations and the thermodynamic equations become identical. 

We chose to separate the system so that one protein species, g1Prot, is in the  subnetwork and the other, g2Prot, is in the bulk. After expanding into protein and DNA species we proceeded to place all DNA species along with g2Prot in the bulk. Because we have performed the expansion around a steady state, the method is more precise when close to the steady state as expected (Fig.~\ref{fig:Expansion}D).
Note that the cross-repressive motif with one binding site is not bistable. It could be made bistable by adding binding sites; with the current approach the behaviour around each steady state would then have to be analysed separately, as we also do in the neural tube application in the Sec.~\ref{sec:app}. More general projection methods can overcome this limitation as detailed in the Discussion, although it is clear that there are limits of principle here. For example, where the dynamical choice of steady state depends crucially on the initial concentration of a bulk species, this effect cannot be captured by a subnetwork description that by definition does not track this species.

Next we analyse the memory functions in our cross-repressive example, which can be thought of as describing the strength of a signal returning from the bulk a specific time after the original signal entered the bulk from the subnetwork (Fig.~\ref{fig:Expansion}E). The amplitude of the leading \emph{linear} memory function -- where the signal is $\delta x_1(t')$ -- is positive. This makes sense as, in the cross repressive motif, a protein is inhibiting its inhibitor, thus promoting its own production. The second order memory function governs the effects of the signal $\delta x_1^2(t')$. It is, in this case, negative and acts as a correction that captures nonlinearities beyond the linear memory. We observe that the decay in time of the memory functions is determined by the decay rate of the bulk protein. This is consistent with the memory function describing a protein repressing its own repressor, hence the effect only lasts as long as the repressor protein is present. Overall, the memory functions provide compact and readable descriptions of how the bulk modifies and influences the activity of the subnetwork.

\subsubsection{Application to neural tube network\label{sec:app}}

Reducing part of a network into memory functions offers a way to simplify and analyse the effect of the factors in this bulk part on the remaining subnetwork. To illustrate this approach we chose to analyse a four gene network involved in the embryonic patterning of the vertebrate neural tube \cite{Dessaud2008,Cohen2013}. The developing neural tube is a well characterised example of developmental pattern formation. In this tissue, the secreted molecule, Sonic-Hedgehog (Shh) forms a ventral to dorsal gradient \cite{Roelink1995}. This in turn generates discrete domains of gene expression that define the progenitors of the distinct neuronal subtypes that comprise spinal cord circuitry (Fig.~\ref{fig:PONI_SSs}A,B). A transcriptional network has been identified that is controlled by graded Shh signaling and is responsible for specifying ventral progenitor domains \cite{Briscoe2000a, Cohen2013}. Genetic and molecular experiments have determined the activity of the components of this network (Fig.~\ref{fig:PONI_SSs}C) and documented the temporal dynamics of pattern formation in vitro and in vivo \cite{Cohen2013, Balaskas2012a, Cohen2014a}. For the three most ventral domains of progenitors, four transcription factors are essential, Nkx2.2, Olig2, Pax6 and Irx3. The most ventral domain (p3 domain) is characterised by high Nkx2.2 and low levels of the other three proteins; the domain adjacent to this (pMN domain) has high levels of Olig2, medium levels of Pax6 and low levels of the other two proteins. Dorsal to this, the p2 domain expresses high levels of Pax6 and Irx3 and low levels of the other two proteins.

Mathematical models have been formulated based on the experimental data and these are able to replicate key aspects of cell patterning \cite{Balaskas2012a,Cohen2014a} (Fig.~\ref{fig:PONI_SSs}D). We take advantage of these to develop a Zwanzig-Mori projection of the system (using equations as described in \cite{Cohen2014a} \& \ref{app:C}). To this end, we chose Nkx2.2 and Olig2 to be the subnetwork species, given that they are receiving direct input from Shh (Fig.~\ref{fig:PONI_SSs}C), and replaced Irx3 and Pax6 with memory functions.

\subsubsection{Linear memory analysis}

We first examined the properties of the linear memory functions. These are the most substantial contributions, close to the specific steady states, of the species in the bulk (Pax6 and Irx3) on the temporal changes in the activity of Nkx2.2 and Olig2. We note that the system is multistable (Fig.~\ref{fig:PONI_SSs}E) and different combinations of steady states are available at different positions along the dorsal ventral axis (Shh gradient). We therefore analysed each possible steady state along the neural tube. In this case, as the bulk species are mutual repressors with the subnetwork species, they form a positive feedback loop with the factors in the subnetwork. As a consequence, each of the linear memory functions is positive and exponentially decaying on the time-scale of protein degradation of the bulk species. Thus the relative contribution of each of the memory functions can be assessed by comparing their amplitude, defined as the memory function value at zero time difference.

\begin{figure}[t]
\centering
\includegraphics[width=1\textwidth, height=0.8\textwidth, keepaspectratio]{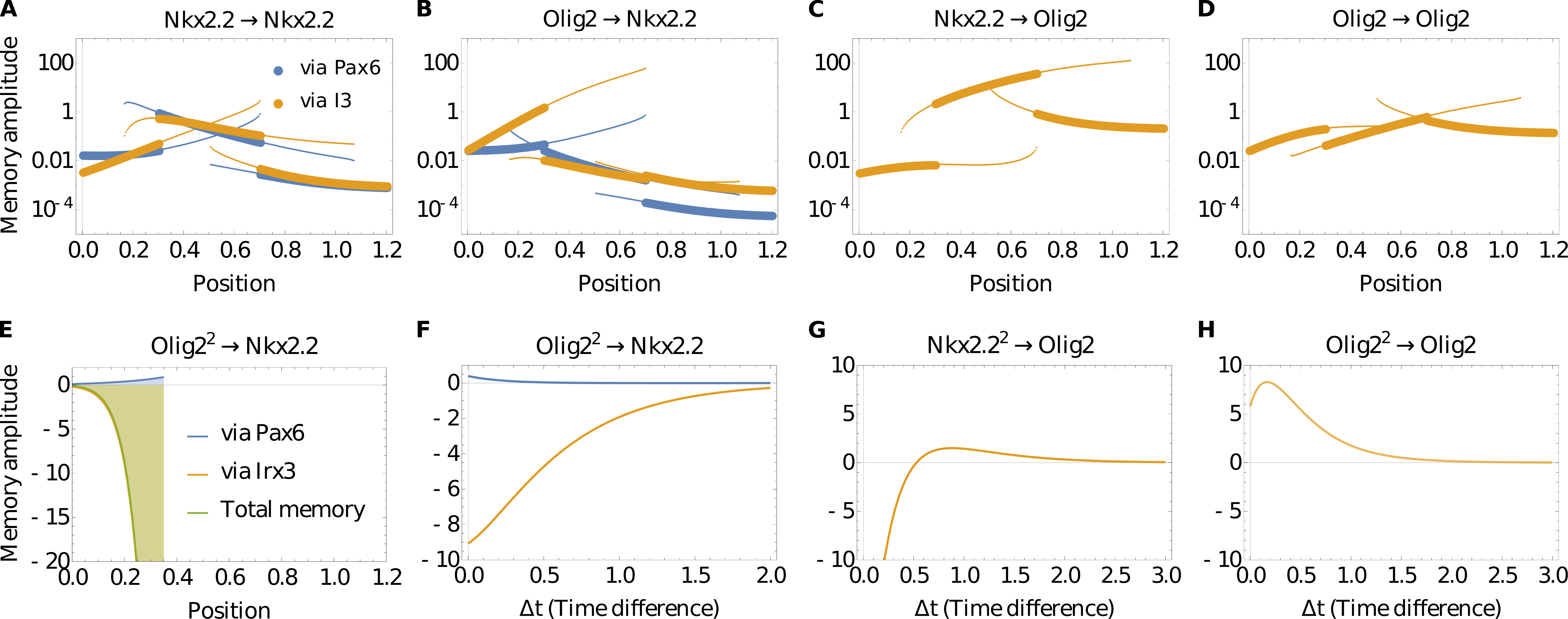}
\caption{{\bf Memory amplitude and temporal dynamics.} (\textbf{A}) Amplitude of memory (memory function at $\Delta t=0$) of Nkx2.2 to itself along the neural tube. There are multiple lines as the analysis was performed at all possible stable steady states. The vertical axis is logarithmic to make the range of amplitudes easier to appreciate. Colours identify the memory amplitude contribution from each of the two possible bulk channels, via Irx3 and Pax6, respectively. Thick lines indicate physiological states, while thin lines indicate states that are not usually observed \textit{in vivo}. (\textbf{B}) Linear memory amplitude of (past) Olig2 on Nkx2.2 along the neural tube. The memory via Pax6 is for the most part below the memory via Irx3 in each pair of corresponding curves. (\textbf{C,D}) Memory amplitudes of Olig2 to Nkx2.2 ({\bf C}) and to itself ({\bf D}). No channel decomposition is performed as Olig2 receives memory only via the Irx3 channel. (\textbf{E}) Nonlinear memory of (past) Olig2 squared on Nkx2.2 in the p3 domain, where dynamics are dominated by Irx3; memory via Pax6 is negligible by comparison. (\textbf{F}) Nonlinear memory function of (past) Olig2 squared on Nkx2.2 from position 0.2 in (\textbf{E}), plotted to show that the relative contribution of the Pax6 channel is small also for all time differences.
(\textbf{G,H}) Nonlinear memory functions of Olig2 to Nkx2.2 ({\bf G}) and itself ({\bf H}) in the p2 domain at position 0.7, exemplifying the potential for nontrivial time dependences (including non-monotonicity and sign changes) in the nonlinear memory functions.}
\label{fig:lMem}
\end{figure}

We calculated the amplitudes of the memory functions and determined the memory effects that Nkx2.2 and Olig2 receive over time. Furthermore, we used the method developed in \cite{Rubin2014} to \emph{decompose} the memory into contributions from memory signals passing through the two different bulk species (Fig.~\ref{fig:lMem}A--D). This allowed us to determine the importance of specific regulatory interactions between transcription factors (TFs) at every position of the neural tube model, and assess their respective contributions. The results indicate that, for the most part, there is only a small memory amplitude of Nkx2.2 to the past of Olig2 passing through Pax6 (Fig.~\ref{fig:lMem}B; note the logarithmic $y$-axis): the memory of (past) Olig2 on Nkx2.2 is largely dominated by memory through Irx3. The only exception is for steady states that are not reached during normal neural tube patterning (lower pair of curves in the neural tube position range $\approx [0.2 \ldots 0.45]$ in Fig.~\ref{fig:lMem}B). These steady states exist because the system is multistable, but the initial conditions that lead to them are incompatible with physiological conditions. We additionally observe in Fig.~\ref{fig:lMem}A that the memory of Nkx2.2 to itself is for the most part shared between Irx3 and Pax6, with no particular TF being dominant. There are again exceptions from this, but only in steady states that are biologically unreasonable. In the case of the memory of Olig2, it can only be influenced by Irx3 as is clear from the structure of the network in Fig.~\ref{fig:PONI_SSs}C. Channel decomposition is therefore unnecessary. We observe that the memory effects of Olig2 to itself are strongest in the p3 domain, and the opposite is true (memory of Nkx2.2 on Olig2 is stronger) in the pMN domain

We observe in Fig.~\ref{fig:PONI_SSs} that the order of magnitude of the memory amplitudes changes substantially with neural tube position. This is a consequence of the system becoming more or less sensitive to concentration fluctuations of a given species within the network. An example of this is Nkx2.2 memory to itself via Pax6 in the pMN domain (Fig.~\ref{fig:lMem}A), where dorsally the lower levels of Shh mean that the binding sites of Nkx2.2 are less occupied by the active form of Gli (Gli is the transcriptional effector downstream of Shh that binds to Nkx2.2 \& Olig2, see \cite{Cohen2014a} for details on how this is implemented). Active Gli sets the rate of production of Nkx2.2 and thus has a direct effect on the amplitude of the memory functions. In addition to this, as one moves dorsally in the pMN domain, the steady state concentration of Pax6 is increasing (Fig.~\ref{fig:PONI_SSs}E, also observed experimentally). This increase causes the system to become insensitive to fluctuations of Pax6 proteins as Pax6 binding sites are typically already saturated. This phenomenon is at work also in the other memory function amplitudes that change across neural tube position as seen in Fig.~\ref{fig:lMem}A-D.

\subsubsection{Nonlinear memory analysis}

We next examined the second order memory functions. These provide corrections to the linear memory terms and encode information about how the bulk nodes drive the system towards a steady state. Quantitative accuracy in capturing the dynamics of the full nonlinear network will also be improved to the extent possible within an expansion around a steady state, but such accuracy is not our main focus -- it is the additional insights gained from the memory functions that we are after.

It is apparent that once memory terms are included, different trajectories in the subnetwork plane \emph{cross}. This is visually most obvious with the full nonlinear memory terms (\ref{fig:nl_effects}~{C,G}) and consistent with the full dynamics (\ref{fig:nl_effects}~{D,H}). The crossings arise from the fact that, in the presence of memory, systems in the same state can follow different trajectories towards a steady state depending on their past. In the full dynamics, the additional information as to which path a system will take is contained in the bulk species, which effectively hold the information on the past of the system. In the case of the memory functions, the memory is not contained in network nodes outside of the subnetwork, but is represented explicitly via the memory terms, with the time difference dependence of the memory functions indicating, for example, the timescale of the memory. This offers a useful abstraction: while it is not easy to visualize the four-dimensional concentration space for the full dynamics, the projection approach  reduces the system to two components with memory of their own past.

In the nonlinear memory functions, we can assess the importance of the link from Olig2 to Pax6 by considering the squared deviations of Olig2: the Pax6 channel contribution to this memory necessarily involves signals being propagated via this Olig2$\to$Pax6 link.  We find that the Pax6 channel makes a very small contribution to the nonlinear memory amplitude in comparison to the Irx3 channel. This is shown in Fig.~\ref{fig:lMem}E for the p3 domain but is true also for all other states. We additionally checked the time-dependence of the memory function: from Fig.~\ref{fig:lMem}F one sees that the relative contribution of the Pax6 channel to the nonlinear memory of (past) Olig2 squared on Nkx2.2 remains small at all time-differences. Together with our results above for linear memory functions, we thus conclude that the repression of Pax6 by Olig2 is dispensable for maintaining an established dorsal-ventral pattern. We plot an example of nonlinear memory functions (Fig.~\ref{fig:lMem}G,H) to illustrate that their dependence on time difference can include non-monotonicities and sign changes that mean memory amplitudes do not tell the full story of the effect of nonlinear memory. The effects of including the memory terms can be confirmed and further visualised by considering the effective drift vector, which contains the rates of change of the concentrations of Nkx2.2 and Olig2. We describe this approach in \ref{app:B}, where we also discuss further how specific contributions of the nonlinear memory affect the dynamics, for example in the approach to the state in the p2 domain.

\begin{figure}[t]
\centering
\includegraphics[width=1\textwidth, height=0.8\textwidth, keepaspectratio]{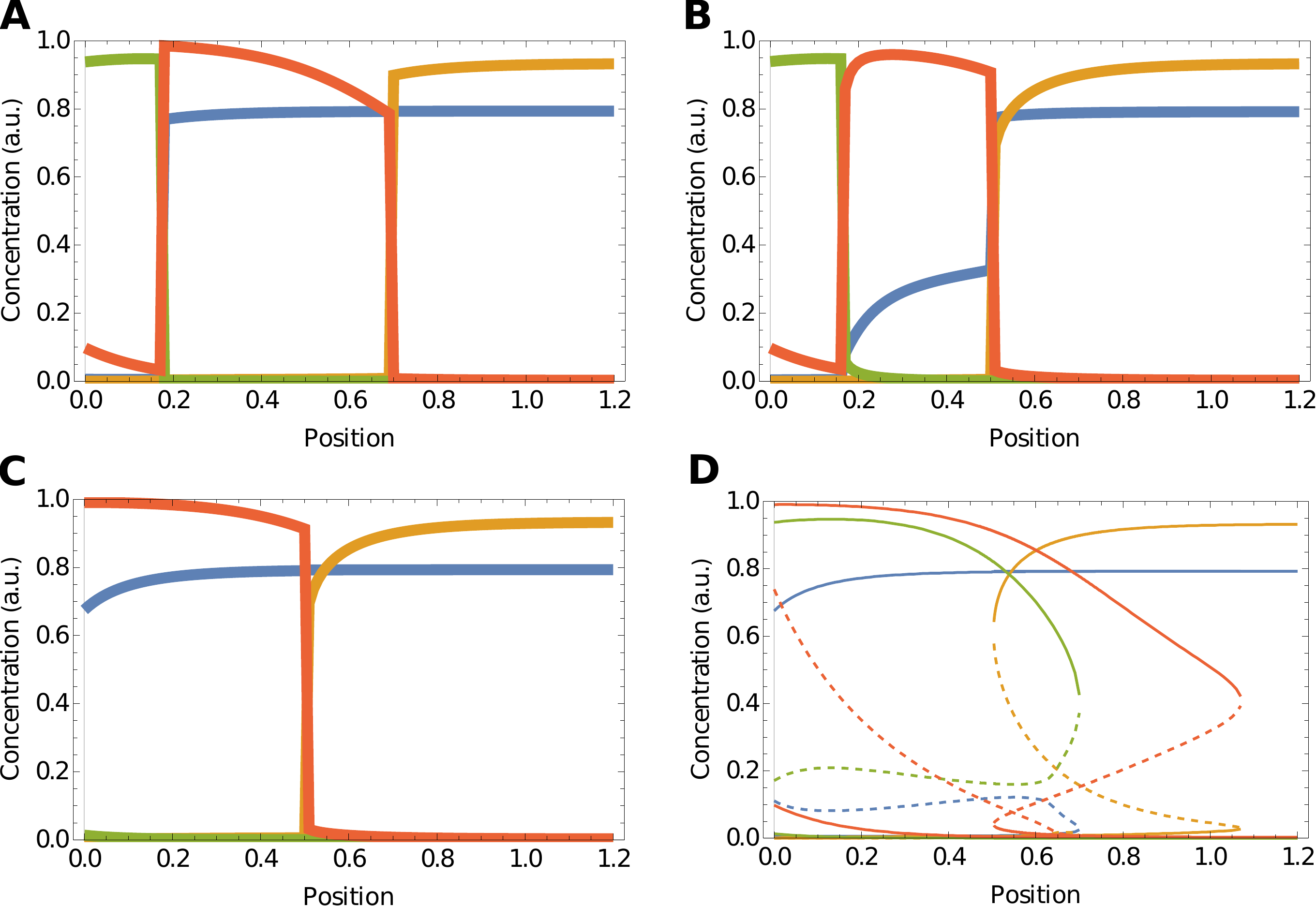}
\caption{{\bf Olig2 repression of Pax6 increases robustness to initial conditions.} (\textbf{A}) Patterning without the repressive link from Olig2 to Pax6, showing steady states reached from standard initial conditions. Compared to the full network, the qualitative domain structure is conserved. (\textbf{B}) Patterning for the full network for initial conditions with high levels of Pax6 and Irx3 is qualitatively identical to low initial levels of Pax6 and Irx3. Initial conditions $[\text{Pax6}]=[\text{Irx3}]=1$. (\textbf{C}) Patterning without the repressive link from Olig2 to Pax6 for initial conditions with high levels of Pax6 and Irx3. The p3 domain is lost as a consequence of the different initial conditions. (\textbf{D}) Bifurcation diagram of the network without Olig2-Pax6 repression. The region in which the pMN state (Olig2 high) is stable has expanded ventrally, thus making the ventralmost region bistable.}
\label{fig:further}
\end{figure}

\subsubsection{Exploration of network properties}
The analysis of both linear and nonlinear memory functions in the previous section suggested that the repression of Pax6 by Olig2 is not critical for the dynamics of the steady states observed in neural tube patterning. We based this on the observation that the amplitude of memory that is transmitted through this link is relatively small compared to the other memory functions, for all biologically relevant steady states.

To test this prediction of the memory function analysis, we removed the repressive link from Olig2 to Pax6 in the full model. Consistent with our prediction of the relative insignificance of the link from Olig2 to Pax6, the resulting steady state concentrations of the simulations are similar to those in the original system (Fig.~\ref{fig:further}A). The only qualitative change is the increased level of Pax6 between neural tube position 0.2 and 0.7, which results from the loss of the repressive link. But the spatial aspects, such as the positional sequence of genes and qualitative concentrations, of the domains are otherwise unaltered.

Since the repressive link between Olig2 and Pax6 has been experimentally documented \cite{Zhou2002,Balaskas2012a,Kutejova2016} but from our analysis is not required for the dynamics around the steady state, we sought to understand what purpose it might serve during development. We performed simulations from varying initial conditions for both WT (Fig.~\ref{fig:further}B) and the system lacking the link (Fig.~\ref{fig:further}C). These show that the p3 domain is reached from only a small range of initial conditions following the removal of the repressive link between Olig2 and Pax6. To rationalise this, we performed a bifurcation analysis (Fig.~\ref{fig:further}D). This indicated that the removal of the Olig2 inhibition of Pax6 markedly increased the range of positions at which the Olig2 steady state is present, thus facilitating its invasion into the Nkx2.2 domain. In particular, under initial conditions of high Pax6 and Irx3, the pMN domain is induced ventrally. In the unperturbed network it is not possible to reach this steady state in the ventralmost part because the system is monostable, allowing only the p3 domain to arise. On the other hand, in the absence of Olig2 inhibition of Pax6, the ventral region is bistable. Hence, depending on the initial conditions a pMN fate can be reached (Fig.~\ref{fig:further}D). Taken together, this analysis suggests that the repressive link from Olig2 to Pax6 contributes to the robustness of the system for patterning the neural tube.

A potential explanation for the seemingly dispensable regulatory interaction between Olig2 and Pax6 arises from considering the normal development of the neural tube. During embryonic neural tube development, neural progenitors are generated by the process of neural induction and this initiates the expression of neural genes including Pax6 and Irx3. At the same time, cells of the notochord (which underlies the ventral midline of the neural tube) begin to secrete Shh. Shh spreads into the neural tube and neural progenitors respond to the signal. The consequence is that in the ventral neural tube neural progenitors begin responding to Shh at intermediate levels of Pax6 and Irx3. A more appropriate initial condition, representative of the in vivo situation, is for Pax6 and Irx3 to be somewhere between 0 and maximum. Moreover, heterogeneity between cells, as well as variation in the exact timing of developmental events along the rostral-caudal axis of the embryo, means that the precise initial conditions -- the levels of Pax6 and Irx3 when cells respond to Shh signalling -- will vary. Hence the robustness provided by the Olig2-Pax6 link may play an important role in ensuring reliable pattern formation in the developing neural tube. 

\section{Discussion\label{sec:disc}}
We have developed a generally applicable method for applying Zwanzig-Mori projections to transcriptional networks that allows the analysis of subnetworks extracted from a larger network. We demonstrated the approach on a simple genetic cross repressive motif, to illustrate the minimal example of a memory function, and on a more complicated, but well-characterised, transcriptional network operating in the ventral part of the vertebrate neural tube. This showed how the method allows the function and importance of specific links within a network to be defined in an intuitive manner. We used these insights to identify structural features of the neural tube network that appear primarily to increase the robustness of the network to initial conditions, rather than maintain its steady state.

The projection technique is straightforward to implement using the methodology described, and can be applied to any GRN following the thermodynamic formalism (including activation, repression, competitive and cooperative binding). Further protein-protein interactions could readily be incorporated as these would be represented by simple first order (unary) or second order (binary) interactions. Thus, in addition to gene regulation, nonlinear protein-protein interactions mediated by enzymes such as those found in signal transduction pathways could be included in the analysis \cite{Rubin2016}. It would also be possible to include Hill functions as long as these possess integer exponents \cite{Thomas2012c}. Note that our treatment relies on mass action kinetics, which implicitly assumes that the different molecular species are well mixed by diffusion processes that are faster than any of the reaction kinetics~\cite{Smith2016}. The size of the matrices involved in constructing the memory functions scales as the square of the number of bulk species, which in our experience means that networks with up to 200 nodes can be investigated without difficulty. Thus, the generality of the method and the ability to implement it algorithmically provide a comprehensive mathematical toolkit to simplify and analyse dynamical systems describing a range of cellular and molecular processes.

In our mathematical treatment we took advantage of an argument of timescale separation for molecular mechanisms operating in transcriptional networks, in which fast processes (DNA-protein interactions) are assumed to be in QSS compared to the slow processes (changes in protein concentration). We also assume the effective concentration of DNA species to be small in comparison to that of protein species. We consider these to be reasonable assumptions because DNA-protein interactions occur at a much faster rate than the changes in protein concentrations produced by transcription and translation~\cite{Garcia2010}. Moreover, there are usually at most four DNA copies of each gene per cell, whereas the number of individual protein molecules is normally considerably higher. The production of individual proteins per DNA copy number, \emph{i.e.}\ the relevant rate constant, can then be large while the resulting relative changes to protein concentration are slow. These assumptions allowed us to construct an extended reaction network involving fast molecular (DNA) species and at most binary reactions. This construction is sufficiently general for it to be applicable to other systems with a timescale separation, as long as they satisfy the above conditions for the relevant concentrations and rate constants.

An important component of this Zwanzig-Mori projection method is that the memory functions describe all the behaviours of a system in the vicinity of a steady state.
Even though this comes at the cost of predictions becoming increasingly inaccurate the further the system is from the fixed point, the information around the steady state is sufficient to distil relevant properties of the network.
We therefore employ the method as a tool for probing the dynamics near a chosen steady state, rather than for tracking dynamics. The benefit is that the resulting memory functions are substantially simpler than the original system. There is no need to perform simulations to analyse memory functions or to test different positions in phase space as a steady state is approached. Moreover, the method can also be applied to an arbitrarily chosen subnetwork of a generic network whereas other model reduction approaches -- as discussed in the Introduction -- do not allow this, and are also more restrictive as they require specific network motifs or functional forms of the dynamics.

The memory functions produced by the Zwanzig-Mori approach allow exploration of the amplitudes and timescales of the interactions between the subnetwork and bulk. This can reveal information that is not otherwise easy to discern from the parameters of the full dynamics or from the functional form of the equations. Each memory function represents the total of the contributions from a given subnetwork component that feeds back to the subnetwork over time after passing through the bulk. This total memory can be decomposed into channels that describe the flow of signals through specific components in the bulk. In this way, the channels that dominate a memory effect can be identified, providing insight into the dynamical mechanisms responsible for achieving and maintaining a steady state. The memory functions thus help to extract features of the dynamics that would not be readily detectable in direct simulations of the time evolution equations.

An example of the type of insight provided by the projection method came from our analysis of the neural tube network. We focused on a subnetwork comprising Nkx2.2 and Olig2. Decomposition of the memory functions suggested that an experimentally documented repression of Pax6 by Olig2 \cite{Zhou2002} plays only a minor role in sustaining the steady state pattern. Consistent with this prediction, simulations in which this regulatory link had been eliminated in the full model showed qualitatively unchanged steady states. We note that removal of the repressive link between Olig2 and Pax6 is distinct from the removal of Olig2 itself \cite{Cohen2013, Balaskas2012a, Cohen2014a}. In the Olig2 mutant, the repressive effect of Olig2 on both Nkx2.2 and Pax6 are eliminated whereas the removal of the Pax6-Olig2 link leaves the repressive effect of Olig2 on Nkx2.2 intact. The memory function analysis thus raised the question of what purpose the Olig2-Pax6 regulation might serve and prompted us to explore the transient system dynamics, specifically the effect of the initial conditions, which are unknown in the \emph{in vivo} system. Simulations in the absence of the Olig2-Pax6 repression, compared to the full model, revealed a marked increase in sensitivity to these initial conditions. Comparison of the bifurcation diagrams of the systems, with and without the Pax6-Olig2 link, indicated that the well-defined Nkx2.2 monostable region (p3 domain) -- which appears at high levels of Shh signalling in the full system -- was replaced by a region of bistability for Nkx2.2 and Olig2 in the absence of the Olig2-Pax6 regulation. In this latter case the steady state induced by high levels of signal therefore depends on the initial concentrations of Pax6 and Irx3. This suggests that, while not necessary to maintain steady states, the Olig2-Pax6 regulation ensures access to the appropriate steady states irrespective of the initial levels of Pax6 and Irx3. As these may differ between cells and at different locations along the rostral-caudal axis of the neural tube, the Olig2-Pax6 regulatory link might make the system less sensitive to these variations. This implies that the primary purpose of Olig2 repression of Pax6 is to increase the robustness of pattern formation.

Alternative implementations of the Zwanzig-Mori projection \cite{Zwanzig1980,Chorin2000,Zwanzig2001} have been developed that do not rely on expanding around a fixed steady state and we are currently adapting these for transcriptional networks. While these alternative methods may provide a solution to the specific limitations of our current Zwanzig-Mori projection method, the nonlinear memory functions they produce are much richer. Further work will be needed to fully understand the more complex information they encode. Projection methods might also provide a way to analyse systems for which only partial information is available. Given a known subnetwork that is not able to fully reproduce experimental observations, the incorporation of memory functions provides a means to explore the possible effects of unknown factors. This could then be used to identify plausible network structures that generate these memory functions. Taken together therefore, the projection approach provides tools to simplify, visualise and explore the behaviour of large networks that would otherwise be difficult to analyse in their entirety.

\section{Acknowledgments}
We are grateful to Katherine Exelby for providing microscopy images.
EHD and PS acknowledge the stimulating research environment provided by the EPSRC Centre for Doctoral Training in Cross-Disciplinary Approaches to Non-Equilibrium Systems (CANES, EP/L015854/1). EHD and JB are supported by the Francis Crick Institute which receives its core funding from Cancer Research UK (FC001051), UK Medical Research Council (FC001051), Wellcome Trust (FC001051 and WT098326MA). RPC is supported by funding from Wellcome Trust (WT098325MA).

\bibliography{library}

\clearpage
\appendix
\numberwithin{equation}{section}
\setcounter{page}{1}
\renewcommand{\thefigure}{S\arabic{figure}}
\setcounter{figure}{0}

\section{Appendix: Justification of heuristic method for quadratic observables \label{app:A}}

In this Appendix we give the details of the projection method with linear and quadratic observables, applied to an expanded network as outlined in Sec.~\ref{sec:NL}. The expanded network is one of unary and binary reactions involving the concentrations of subnetwork proteins, $\x{s}$, of bulk proteins, $\x{b}$, and of DNA species, $\bx^\a$. The mass action kinetics for this network can be put into the form of an
%into a single vector of slow variables $\bx^\l$ find the form of the effective nonlinear $\LMb{}$ matrix and how that relates to the extended mass action form. We start from an 
$\LMb{}$-matrix, defined for linear and quadratic observables as in (\ref{eq:Ldef}). We partition this into blocks according to
\begin{align}
	\LMb{}=
	\begin{pmatrix}	
	\LMb{S,S}&\vline&\LMb{S,B}\\
	\hline\rule{0pt}{3ex}
	\LMb{B,S}&\vline&\LMb{B,B}
	\end{pmatrix}&=
	\begin{pmatrix}
	\LMbn{s}{s}&\vline&\LMbn{s}{b}&\LMbn{s}{a}\\
	\hline\rule{0pt}{3ex}
	\LMbn{b}{s}&\vline&\LMbn{b}{b}&\LMbn{b}{a}\\
	\LMbn{a}{s}&\vline&\LMbn{a}{b}&\LMbn{a}{a}
	\end{pmatrix}\label{matL_PD_NL}
\end{align}	
Here \~{s} contains the ``subnetwork only'' observables \{s\} (linear) and \{ss\} (quadratic, like $\delta x_s \delta x_{s'}$), while \{\~{b}\} collects the slow bulk observables \{b,~sb,~bb\}. The fast bulk observables are gathered in \{\~{a}\}, which contains \{a,~sa,~ba,~aa\}. Note that with this partitioning of observables we have allocated all fast (DNA) species to the bulk. This is different from the approach in \cite{Rubin2016} where some fast (enzyme) species were retained in the subnetwork in order to retain more of the nonlinearities. In our case one could similarly keep in the subnetwork those DNA species that produce subnetwork proteins, but it turns out that this makes the final elimination of fast variables rather intricate and so we leave this as an option to pursue in future work.

For our GRN equations, subnetwork and bulk protein species do not interact, so the blocks $\LMbn{s}{b}$ and $\LMbn{b}{s}$ are in fact zero. This restriction is not required for our treatment, however, and direct protein-protein interactions could be included in the formalism without modification. As in the case of the linearised dynamics (Sec.~\ref{sec:equivLin}), only the third column of (\ref{matL_PD_NL}) is fast, \emph{i.e.}\ has entries proportional to $\gamma$ (plus subleading terms of order unity arising from the time derivatives of slow-fast product observables such as {ba}).

%\subsection{Aim}
From (\ref{matL_PD_NL}) one can obtain the rate matrix and memory functions for the projected subnetwork equations, for any finite $\gamma$. The limiting values of these quantities for $\gamma\to\infty$ can then be found from a matrix $\bL\eff$ for only the slow (protein) observables. Our aim in this Appendix is to show that this $\bL\eff$ is identical to the analogous matrix that one obtains by directly expanding the original slow (GRN) equations to second order in the protein concentrations. This then justifies the heuristic method of constructing the nonlinear rate matrix and memory functions described in Sec.~\ref{sec:NL}.

\subsection{Generic form of notation}

It will be useful to write the full time evolution equations for the expanded network in the generic form
%We discuss how the matrix $\bL\eff$ for projection with linear and quadratic observables can be obtained. It is derived by taking a fast rate ($\gamma\to\infty$) limit of the projection results for an expanded network where concentrations evolve in time according to 
\begin{align}
\partial_t \bx^\l &= \bR^\l(\bx^\l,\bx^\a)\\
\gamma^{-1} \partial_t \bx^\a &= \bR^\a(\bx^\l,\bx^\a)
\end{align}
These are equations (\ref{eq:oriP}, \ref{eq:meDNAG}) from the main text, or more generically (\ref{eq:TD_QSS}), but we have grouped together the subnetwork and bulk concentration vectors $\x{s}$ and $\x{b}$ into a single vector of slow variables $\bx^\l$ to keep the notation for the following discussion compact. The vector $\bx^\a$ contains the fast variables, which in the GRN context are concentrations of DNA conformations, while $\gamma$ is a fast rate parameter as before.
In the limit of large $\gamma$, the fast variables are always in QSS with the slow ones so that the expanded network reduces to the thermodynamic state ensemble (in the GRN case) dynamics
\begin{align}
\partial_t \bx^\l &= \bR^\l(\bx^\l,\bx^\a)\\
0 &= \bR^\a(\bx^\l,\bx^\a)
\end{align}
where the second equation implicitly determines $\bx^\a$ as a function of $\bx^\l$. (In the main text we marked this QSS value by an asterisk; we omit this here for notational simplicity.) This is the generic form of equations (\ref{eq:oriP},\ref{eq:DNAw}) in the main text. 

The matrix $\LMb{}$ is obtained in the above generic setting by expanding around a fixed point to second order to write the equations of motion as
\begin{eqnarray}
\partial_t \bx^\l{}\T &=& \bx^\l{}\T \bL^{\l\l} + \bx^\a{}\T \bL^{\a\l} +
\bx^{\l\l}{}\T \bL^{\l\l,\l} + \bx^{\l\a}{}\T \bL^{\l\a,\l} + \bx^{\a\a}{}\T \bL^{\a\a,\l} 
\label{eq:bxlproj}
\\
\gamma^{-1} \partial_t \bx^\a{}\T &=& \bx^\l{}\T \bL^{\l\a} + \bx^\a{}\T \bL^{\a\a} +
\bx^{\l\l}{}\T \bL^{\l\l,\a} + \bx^{\l\a}{}\T \bL^{\l\a,\a} + \bx^{\a\a}{}\T \bL^{\a\a,\a} 
\label{eq:bxaproj}
\end{eqnarray}
All $\bx$ appearing here and below are deviations $\delta \bx$ from steady state; we drop the $\delta$ to lighten the notation.
The $\bx^{\l\l}$ etc are product variables -- we assume the indices are ordered to avoid duplicate observables -- and the $\bL$ matrices contain the appropriate derivatives of the ``drift'' functions $R$ at the fixed point, \emph{e.g.}\ $\bL^{\l\l,\a}$ has elements $L_{ll',a}= \partial_{x_l}\partial_{x_{l'}}R_a$ for $l<l'$ and $L_{ll,a}= (1/2)\partial_{x_l}\partial_{x_{l}}R_a$ when the two indices are equal. From the above equations then follow the evolution equations for the slow product variables $\bx^{\l\l}$ and the fast products $\bx^{\l\a}$ and $\bx^{\a\a}$; see (\ref{xla_eqn}, \ref{eq:xaa_condition}) below. From the product rule these equations only involve product variables on the r.h.s.; third order terms are in principle present but discarded within the second order expansion.
Collecting all variables into a vector $\bm{z}$ that concatenates $\bx^{\l}$, $\bx^{\l\l}$, $\bx^{\a}$, $\bx^{\l\a}$, $\bx^{\aa}$ gives the time evolution equation in the form $\partial_t \bm{z}\T = \bm{z}\T \LMb{}$, where the matrix $\LMb{}$ has the block form (\ref{matL_PD_NL}) if one restores the split of slow observables into subnetwork and bulk.

%\subsubsection{Statement of the equivalence}

%In this Appendix we show that $\bL\eff$ can equivalently be obtained directly, by expanding the slow equations to quadratic order in $\bx^\l$. This is convenient in cases where these slow equations are the starting point of the analysis anyway and the fast species are necessary only to bring the system into the form of a network of unary and binary reactions to which the projection method of~\cite{Rubin2014} can be applied.

\subsection{Heuristic method}

%\subsubsection{Direct expansion of slow equations}
The heuristic method involves a direct expansion of the slow equations. To obtain the general form of this, one writes the dynamical equations as
\begin{eqnarray}
\partial_t  \bx^\l{}\T &=& \bx^\l{}\T \bL^{\l\l} + \bx^\a{}\T \bL^{\a\l} +
(\bx^\l\circ\bx^\l)\T \bL^{\l\l,\l} + 
(\bx^\l\circ\bx^\a)\T \bL^{\l\a,\l} + 
(\bx^\a\circ\bx^\a)\T \bL^{\a\a,\l}
\label{eq:bxl}
\\
\gamma^{-1} \partial_t \bx^\a{}\T &=& \bx^\l{}\T \bL^{\l\a} + \bx^\a{}\T \bL^{\a\a} +
(\bx^\l\circ\bx^\l)\T \bL^{\l\l,\a} + 
(\bx^\l\circ\bx^\a)\T \bL^{\l\a,\a} + 
(\bx^\a\circ\bx^\a)\T \bL^{\a\a,\a}
\label{eq:bxa}
\end{eqnarray}
where the ``circle'' product indicates the actual products of the regular linear observables, with the same index ordering as in the projection approach (so that all the $\bL$-matrices are as before). One now needs to determine $\bx^\a$ by setting the r.h.s.\ of (\ref{eq:bxa}) to zero, and substitute the result into (\ref{eq:bxl}). As we are only expanding to second order in $\bx^\l$, it is enough also to obtain $\bx^\a$ to this order.
Starting with the first order of (\ref{eq:bxa}) one obtains
\begin{equation}
\bx^\a{}\T = -\bx^\l{}\T \bL^{\l\a}(\bL^{\a\a})^{-1} \equiv \bx^\a_0{}\T
\end{equation}
where $\bx^\a_0$ will be a convenient shorthand. (Note that the coefficient matrix $-\bL^{\l\a}(\bL^{\a\a})^{-1}$ is the one we worked out in the main text below (\ref{eq:linearised_qss}), in a slightly more pedestrian fashion.)
All \emph{second} order terms in (\ref{eq:bxa}) can now be evaluated to the required accuracy by replacing $\bx^\a$ by $\bx^\a_0$. Solving for $\bx^\a$ then gives
\begin{equation}
\bx^\a{}\T = \bx^\a_0{}\T - \left[
(\bx^\l\circ\bx^\l)\T \bL^{\l\l,\a} + 
(\bx^\l\circ\bx^\a_0)\T \bL^{\l\a,\a} + 
(\bx^\a_0\circ\bx^\a_0)\T \bL^{\a\a,\a}\right](\bL^{\a\a})^{-1}
\end{equation}
Inserting into (\ref{eq:bxl}) and dropping terms of higher than quadratic order gives the required expansion of the slow equations,
\begin{eqnarray}
\partial_t \bx^\l{}\T &=& \bx^\l{}\T \bL^{\l\l} + \bx^\a_0{}\T \bL^{\a\l} +
(\bx^\l\circ\bx^\l)\T \bL^{\l\l,\l} + 
(\bx^\l\circ\bx^\a_0)\T \bL^{\l\a,\l} + 
(\bx^\a_0\circ\bx^\a_0)\T \bL^{\a\a,\l}
\nonumber\\
&&{}-\left[
(\bx^\l\circ\bx^\l)\T \bL^{\l\l,\a} + 
(\bx^\l\circ\bx^\a_0)\T \bL^{\l\a,\a} + 
(\bx^\a_0\circ\bx^\a_0)\T \bL^{\a\a,\a}\right](\bL^{\a\a})^{-1}\bL^{\a\l}
\end{eqnarray}
This then determines the $\LMb{}$-matrix for the slow variables from which rate matrix and memory functions are determined in the heuristic approach.

\subsection{Expanded network approach}

As explained in the main text based on the reasoning in \cite{Rubin2016}, if one writes down expressions for the rate matrix and memory functions from the full matrix $\LMb{}$ and then takes the fast rate limit $\gamma\to\infty$,  the resulting rate matrix and (slow) memory function can be found from a matrix $\bL\eff$ describing only the dynamics of the slow variables $\bx^\l$ and $\bx^{\l\l}$. This $\bL\eff$ is obtained by eliminating the fast variables $\bx^\a$, $\bx^{\l\a}$ and $\bx^{\a\a}$ using the condition that they are in QSS. These conditions are, within the projection approach, \emph{linear} equations because product observables are treated as independent from linear observables.
%The projection approach has the same starting point as the direct expansion, with (\ref{eq:bxlproj},\ref{eq:bxaproj}) being equivalent to (\ref{eq:bxl},\ref{eq:bxa}). The difference is that we have separate conditions from which to eliminate the fast product variables, via the QSS requirement for these variables. So 

What we then need to demonstrate is that this distinct elimination assigns to $\bx^{\l\a}$ and $\bx^{\a\a}$ the same values as the direct expansion approach, namely $(\bx^\l\circ\bx^\a_0)$ and $(\bx^\a_0\circ\bx^\a_0)$. Once this is shown, it follows that the linear fast variables $\bx^\a$ are eliminated in the same way in the two approaches, because the same quadratic fast variables are substituted into the relevant equations (\ref{eq:bxaproj},\ref{eq:bxa}). Thus \emph{all} fast variables are eliminated in the same way from the time evolution equation for the slow variables, to the quadratic order we consider here. As the slow $\LMb{}$-matrix in the heuristic approach and the $\bL\eff$ in the expanded networm method are both obtained from this time evolution equation for the slow variables, they are therefore equal as we want to show.

\subsubsection{Elimination of $\bx^{\l\a}$}

Using the product rule of differentiation, the equations of motion for the $\bx^{\l\a}$ are
\begin{equation}
\partial_t \bx^{\l\a}{}\T = 
\gamma ((\bx^\l{}\T\circ \bx^\l{}\T \bL^{\l\a})) +
\gamma ((\bx^\l{}\T\circ \bx^\a{}\T \bL^{\a\a})) +
((\bx^\l{}\T \bL^{\l\l}\circ \bx^\a{}\T)) +
((\bx^\a{}\T \bL^{\a\l}\circ \bx^\a{}\T))
\label{xla_eqn}
\end{equation}
Here the double brackets on the right indicate that after the circle products are taken the real products have to be replaced by product variables, to remain within the projection framework.

We want to eliminate the $\bx^{\l\a}$ from the condition that the r.h.s.\ vanishes. Fortunately for large $\gamma$ the first two terms, which stem from the time evolution of $\bx^\a$, dominate; the last two become negligible in comparison. 
Thus one has to solve
\begin{equation}
((\bx^\l{}\T\circ \bx^\l{}\T \bL^{\l\a})) +
((\bx^\l{}\T\circ \bx^\a{}\T \bL^{\a\a})) = 0
\label{eq:xal_condition}
\end{equation}
From the structure of this one sees that the $\bx^\l$ only act as ``spectators'', while considering the second factors one has to solve the same equation as at linear order. The solution is therefore expected to be $\bx^{\l\a}{}\T=((\bx^\l{}\T\circ\bx^\a_0{}\T))$ as we want to show.

To see this in more detail we write out (\ref{eq:xal_condition}) in components:
\begin{equation}
\sum_{l'} x_{(ll')} L_{l'a} +
\sum_{a'} x_{la'} L_{a'a}= 0
\label{eq:xal_components}
\end{equation}
Here we have written $x_{(ll')}$ to indicate that the indices are to be taken as ordered, \emph{i.e.}\ $x_{(ll')}=x_{ll'}$ if $l\leq l'$ and $x_{(ll')}=x_{l'l}$ otherwise. The proposed solution is
\begin{equation}
x_{la} = ((\bx^\l{}\T\circ\bx^\a_0{}\T))_{la} = -\sum_{l'a'}
x_{(ll')} L_{l'a'} (\bL^{\a\a})^{-1}_{a'a}
\label{eq:xla_solution}
\end{equation}
This does solve (\ref{eq:xal_components}) because
\begin{equation}
\sum_{a''} x_{la''} L_{a''a} =
-
\sum_{l'a'a''}
x_{(ll')} L_{l'a'} (\bL^{\a\a})^{-1}_{a'a''}
 L_{a''a} = 
-\sum_{l'} x_{(ll')} L_{l'a} 
\end{equation}

\subsection{Elimination of $\bx^{\a\a}$}

We proceed again using the product rule of differentiation, which gives as the equations of motion for the $\bx^{\a\a}$
\begin{equation}
\partial_t \bx^{\a\a}{}\T = 
\gamma ((\bx^\l{}\T \bL^{\l\a}\circ \bx^\a{}\T)) +
\gamma((\bx^\a{}\T \bL^{\a\a}\circ \bx^\a{}\T))+
\gamma ((\bx^\a{}\T\circ \bx^\l{}\T \bL^{\l\a})) +
\gamma ((\bx^\a{}\T\circ \bx^\a{}\T \bL^{\a\a}))
\label{eq:xaa_condition}
\end{equation}
Here all terms contribute for large $\gamma$, but one sees that in the first two the right factor of $\bx^\a$ is again a ``spectator'' and similarly with the left factor for the last two terms. Accordingly one can show that the proposed solution, which is
$\bx^{\a\a}{}\T=((\bx^\a_0{}\T\circ\bx^\a_0{}\T))$, 
ensures that each pair of terms vanishes separately.

For explicit calculation it is again useful to write out components. The $aa'$ component of the first two terms of
(\ref{eq:xaa_condition}), without the overall factor of $\gamma$, reads
\begin{equation}
\sum_{l'} x_{l'a'} L_{l'a} +
\sum_{a''} x_{(a'a'')} L_{a''a}
\label{eq:xaa_components}
\end{equation}
The proposed solution is
\begin{equation}
x_{aa'} = ((\bx^\a_0{}\T\circ\bx^\a_0{}\T))_{aa'} = \sum_{l_1 l_2 a_1 a_2}
x_{(l_1l_2)} L_{l_1 a_1} (\bL^{\a\a})^{-1}_{a_1a}
L_{l_2 a_2} (\bL^{\a\a})^{-1}_{a_2 a'}
\label{eq:xaa_solution}
\end{equation}
Substituting this and the solution (\ref{eq:xla_solution}) for $x_{la}$ turns (\ref{eq:xaa_components}) into
\begin{equation}
-\sum_{l'l''a''}
x_{(l'l'')} L_{l''a''} (\bL^{\a\a})^{-1}_{a''a'}
 L_{l'a} +
\sum_{a'' l_1 l_2 a_1 a_2}
x_{(l_1l_2)} L_{l_1 a_1} (\bL^{\a\a})^{-1}_{a_1a'}
L_{l_2 a_2} (\bL^{\a\a})^{-1}_{a_2 a''}
L_{a''a}
\end{equation}
The last factors in the second term again cancel, reducing it to
\begin{equation}
\sum_{l_1 l_2 a_1}
x_{(l_1l_2)} L_{l_1 a_1} (\bL^{\a\a})^{-1}_{a_1a'}
L_{l_2 a}
\end{equation}
After a relabelling of summation indices this is identical to the first term. This means that (\ref{eq:xaa_components}) vanishes, \emph{i.e.}\ the first two terms on the r.h.s.\ of (\ref{eq:xaa_condition}) cancel. Similarly the last two terms vanish, showing that (\ref{eq:xaa_solution}) is the correct QSS assignment of the $\bx^{\a\a}$.

\subsubsection{Equations of motion for $\bx^{\l\l}$}

Above we have shown that the direct and the projection elimination procedures give the same time evolution equation for $\bx^\l$. The same can then also be checked straightforwardly for the product variables $(\bx^\l\circ \bx^\l)$ and their projection analogues $\bx^{\l\l}$. By analogy with (\ref{xla_eqn}), the latter evolve in time according to
\begin{equation}
\partial_t \bx^{\l\l}{}\T = 
((\bx^\l{}\T\circ \bx^\l{}\T \bL^{\l\l})) +
((\bx^\l{}\T\circ \bx^\a{}\T \bL^{\a\l})) +
((\bx^\l{}\T \bL^{\l\l}\circ \bx^\l{}\T)) +
((\bx^\a{}\T \bL^{\a\l}\circ \bx^\l{}\T))
\label{xll_eqn}
\end{equation}
The real products obey the same equation, just written differently:
\begin{equation}
\partial_t (\bx^\l\circ\bx^\l)\T = 
(\bx^\l{}\T\circ \bx^\l{}\T \bL^{\l\l}) +
(\bx^\l{}\T\circ \bx^\a{}\T \bL^{\a\l}) +
(\bx^\l{}\T \bL^{\l\l}\circ \bx^\l{}\T) +
(\bx^\a{}\T \bL^{\a\l}\circ \bx^\l{}\T)
\end{equation}
From both, the fast products $(\bx^\l\circ\bx^\a)$ (respectively $\bx^{\l\a}$) and $(\bx^\a\circ\bx^\a)$ (respectively $\bx^{\a\a}$) then need to be eliminated. As we have already established that these eliminations are identical, also the resulting equations for the ll-product variables must be identical.

\clearpage
\setcounter{page}{1}

\section{Appendix: Further analysis of memory effects \label{app:B}}

In this Appendix we give additional detail of the wide variety of effects that memory functions contribute to the behaviour of a system. Specifically, we explore the effect of the contributions from nonlinear memory terms and provide an alternative visualisation using a first order time expansion of the effective drift of the system.

\subsection{Nonlinear memory effects}

\begin{figure}[h]
\centering
\includegraphics[width=1\textwidth, height=0.6\textwidth, keepaspectratio]{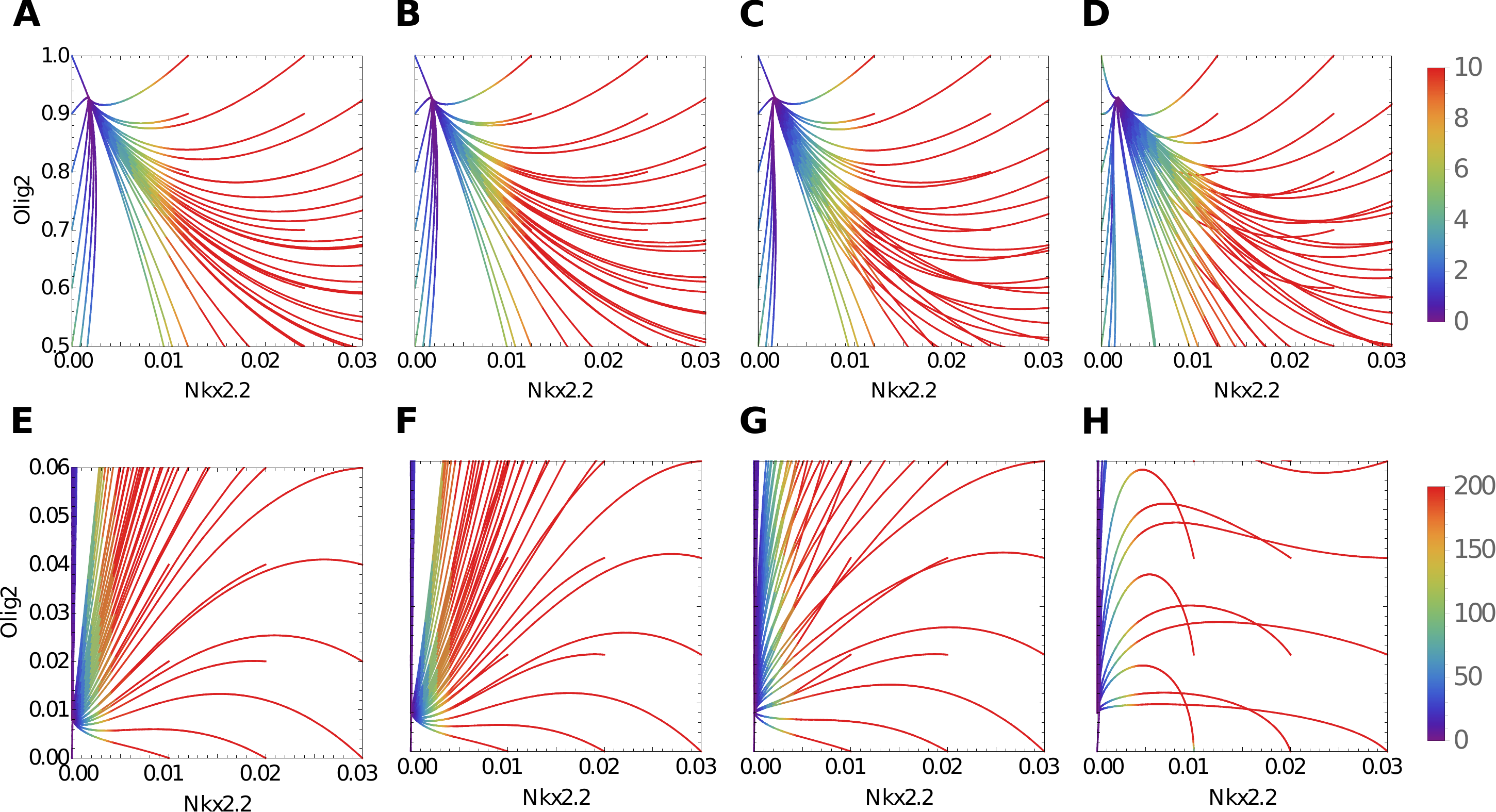}\\
\caption{{\bf Trajectories approaching a steady state, with and without memory.} Trajectories approaching a high Olig2 state in the pMN domain from a range of initial conditions, with no memory (\textbf{A}), with only linear memory (\textbf{B}), with linear and nonlinear memory (\textbf{C}), and the full dynamics (\textbf{D}). Trajectories approaching a low Olig2, low Nkx2.2 state in the p2 domain from different initial conditions, with no memory (\textbf{E}), with only linear memory (\textbf{F}), with linear and nonlinear memory (\textbf{G}), and the full dynamics (\textbf{H}). 
All figures are parametric plots, showing Nkx2.2 and Olig2 concentrations on the $x$- and $y$-axis, respectively, with time as curve parameter. The trajectories are coloured to represent the norm of the drift vector of the system as indicated by the colour scale; high values indicate the system is evolving quickly while the opposite is true for low values. Scalebars on the right apply to the corresponding row. Path crossing can be seen in (\textbf{C}--\textbf{D}) \& (\textbf{G}--\textbf{H}), illustrating the importance of nonlinear memory terms to reproduce qualitative features of the full thermodynamic equations.}
\label{fig:nl_effects}
\end{figure}

\begin{figure}[h]
\centering
\includegraphics[width=1\textwidth, height=0.6\textwidth, keepaspectratio]{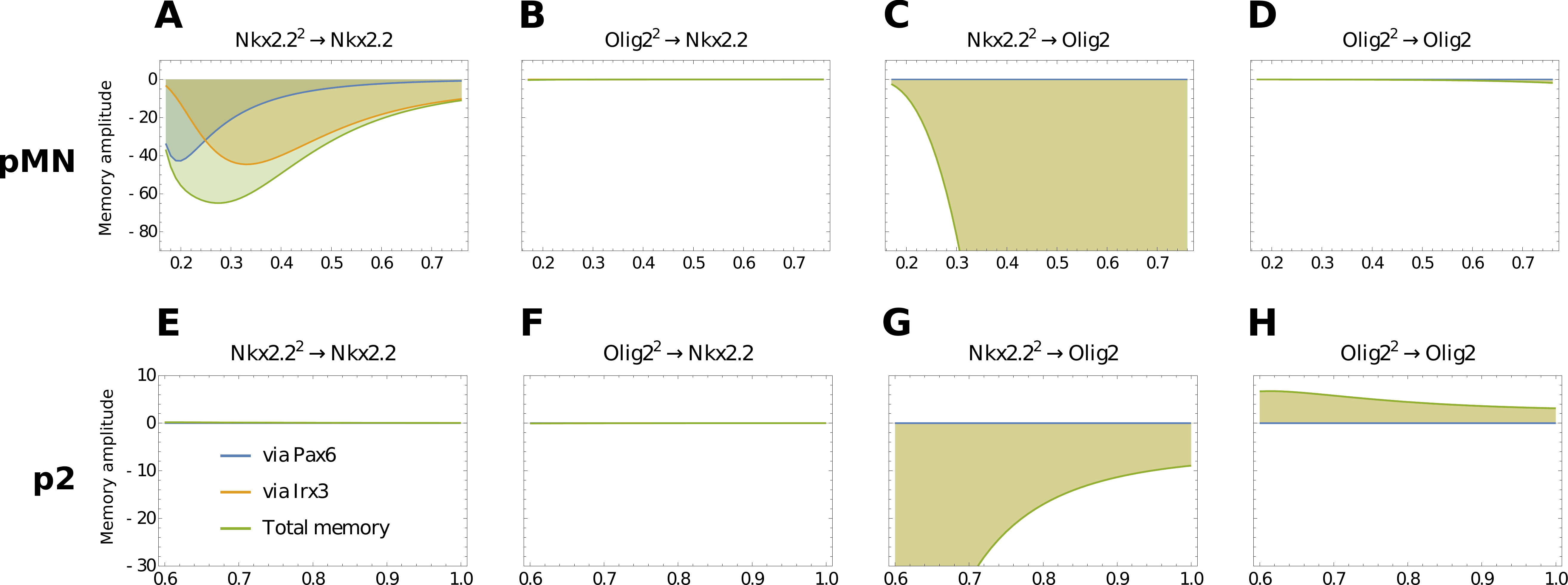}\\
\caption{{\bf Nonlinear memory amplitudes.} (\textbf{A}-\textbf{D}) Nonlinear memory amplitudes across neural tube positions within the pMN domain. The plot titles indicate the type of memory, \emph{e.g.}\ (\textbf{C}) shows memory of (past) Nkx2.2 squared fluctuation on Olig2. Memory effects to past Olig2-fluctuations in the pMN domain are clearly negligible. (\textbf{E}-\textbf{H}) Nonlinear memory amplitudes across neural tube positions within the p2 domain. Nkx2.2 visibly receives very little quadratic memory influence in this domain. The $x$-axis represents neural tube position in all plots. Blue and yellow lines indicate the decomposition into Pax6 and Irx3 channels, while green lines indicate the total memory.}
\label{fig:nl_expl}
\end{figure}

In Fig.~\ref{fig:nl_effects} we show trajectory plots comparing the predictions without memory (left column) to those with linear and linear+quadratic memory (middle columns) and to the dynamics of the full network.
Focussing directly onto the description with all memory terms included, we can gain insight into what drives the distinct dynamics of Fig.~\ref{fig:nl_effects}C,G by analysing the respective nonlinear memory functions, specifically their amplitude (value at $\Delta t=0$) as shown in Fig.~\ref{fig:nl_expl}A-D. We note that in most cases the memory contributions arise via Irx3. (In the case of memory of Olig2 this is the only possibility in any case.) A clear exception to this is the case of nonlinear Nkx2.2 memory to itself (Fig.~\ref{fig:nl_expl}A) where the memory is dominated ventrally by the Pax6 channel and dorsally by the Irx3 channel. These memory functions also change with time difference, but it turns out that they decay on similar timescales so that the memory amplitude gives a reasonable characterization of their strength. We observe that Olig2 does not produce memory effects that are substantial enough to change the dynamics (Fig.~\ref{fig:nl_expl}B,D): the dominant nonlinear memory effects come from Nkx2.2 (Fig.~\ref{fig:nl_expl}A,C). We see further that Nkx2.2 represses both itself (negative nonlinear memory amplitude, Fig.~\ref{fig:nl_expl}A) and, even more strongly, Olig2 (Fig.~\ref{fig:nl_expl}C).
These two terms combined mean that at high levels of Nkx2.2, the memory functions will impede Olig2 from increasing as quickly as it would in cases without nonlinear memory, while simultaneously decreasing the concentration of Nkx2.2. Both effects are consistent with what we observe in Fig.~\ref{fig:nl_effects}C, D.

In the p2 domain the memoryless and linear memory predictions for trajectories are quite close, with levels of Nkx2.2 and Olig2 decreasing on similar timescales towards the steady state (Fig.~\ref{fig:nl_effects}E--H). Nonetheless, plots showing the norm of the effective drift indicate the trend that memory pushes the system more quickly towards low Nkx2.2 (Fig.~\ref{fig:supp}D,E, see below). Including the nonlinear memory then further enhances this tendency: the levels of Nkx2.2 drop more rapidly than Olig2 on the approach to the steady state (Fig.~\ref{fig:nl_effects}G,H \& Fig.~\ref{fig:supp}F). The behaviour is not identical to but mimics that of the full thermodynamic model:  the nonlinear memory effects contribute qualitative information about the behaviour of the system further away from the steady state. This points to memory terms being important for first reducing levels of Nkx2.2 before reducing Olig2 levels.

Understanding why this is the case requires a more detailed analysis of the memory effects. The memory amplitudes show that Nkx2.2 is not subject to substantial memory effects (Fig.~\ref{fig:nl_expl}E,F) and the largest memory effects act on Olig2 (Fig.~\ref{fig:nl_expl}G,H). At first glance,  the latter memory amplitudes suggests that Nkx2.2 substantially represses Olig2 while Olig2 activates itself to a lesser degree. However, on closer inspection of the temporal dependence of the memory functions, the repression by Nkx2.2 is a short pulse that for larger $\Delta t$ turns into an activation of Olig2 (Fig.~\ref{fig:nl_effects}F). Meanwhile, the Olig2 self-activation is a sustained signal that lasts much longer than the repression by Nkx2.2. When taken together, these two memory terms (combined with cross terms, not shown) promote activation of Olig2 from high levels of Nkx2.2 even after a substantial time difference has elapsed, of the order of the decay time of the linear memory functions.
This is consistent with what we observe in Fig.~\ref{fig:nl_effects}G,H: the initial high levels of Nkx2.2 lead to Olig2 levels being sustained via the memory, while Nkx2.2 decays in a way almost unaffected by memory. Eventually Olig2 also decays, once enough time has passed for the nonlinear memory effects to fade away.

\subsection{Calculation of effective drift in the presence of memory}
It is useful to be able to visualize the effects of memory terms in the projected equations in terms of an effective drift. This is possible at least perturbatively for short times, as we now show. We illustrate the method for the linearized dynamics, where the projected equation (\ref{eq:lin_proj}) without the random force read in vector form:
\begin{align}
	\partial_t \bm{x}^{\rm T}=\bm{x}(t)^{\rm T}\bm{\Omega}+\int_0^tdt'\bm{x}(t)^{\rm T}\bm{M}(t-t')
\end{align}
For small times $t$, the memory function can be treated as approximately constant so that one obtains in an expansion to first order in $t$:
\begin{align}
	\int_0^tdt'\bm{x}(t)^{\rm T}\bm{M}(t-t')\simeq t\,\bm{x}(t)^{\rm T}\bm{M}(0)
\end{align}
Combining with the rate matrix term gives the effective drift
\begin{align}
	\partial_t \bm{x}^{\rm T}=\bm{x}(t)^{\rm T}[\bm{\Omega}+t \bm{M}(0)]\label{eq:EffDrift}
\end{align}
On the right hand side one has a function of the current concentrations only, as announced. Given that this is a linear expansion in $t$ we refrain from exploring the system too far away from $t=0$. Higher order expansions in $t$ could be performed yet the first order is enough to demonstrate the qualitative contributions of the memory terms. 

\begin{figure}[h]
\centering
\includegraphics[width=0.9\textwidth, height=0.8\textwidth, keepaspectratio]{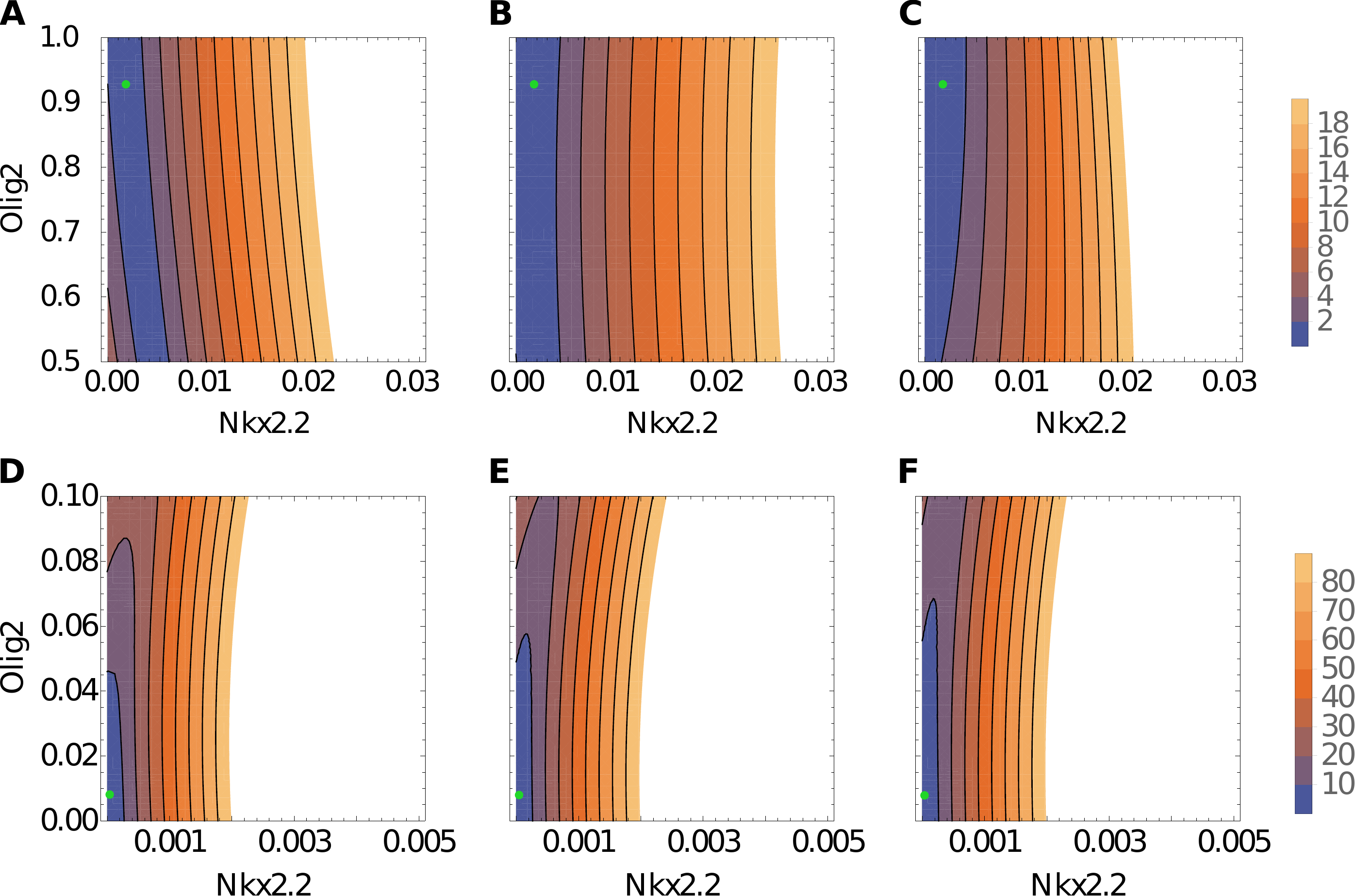}
\caption{Colour map and contour plots of the norm of the effective drift, in conditions identical to those in Fig.~\ref{fig:nl_effects}A-C for the first row (pMN domain) and Fig.~\ref{fig:nl_effects}E-G (p2 domain) for the second row. (\textbf{A},\textbf{D}) Memoryless drift, (\textbf{B},\textbf{E}) drift with linear memory, (\textbf{C},\textbf{F}) drift with nonlinear memory. The scalebar on the right applies to each entire row. Effective drift norms have been calculated for time $t=0.8$.}
\label{fig:supp}
\end{figure}

We illustrate the approach in Fig.~\ref{fig:supp}A-F with contour plots of the norm of the effective drift vector on the r.h.s.\ of (\ref{eq:EffDrift}), for time $t=0.8$.
Bearing in mind that the system will pass quickly through regions where the drift is high and spend most of its time in regions where it is low, these plots confirm the effects seen in the trajectory plots: inclusion of the memory terms causes the low-drift region to shift to lower Nkx2.2 concentrations, \emph{i.e.}\ the system will more rapidly reduce Nkx2.2 and then spend more time increasing Olig2 at small Nkx2.2.

\clearpage
\setcounter{page}{1}

\section{Appendix: Model details\label{app:C}}
The full network equations for the neural tube application in Sec.~\ref{sec:app} are of the form (\ref{eq:oriP}, \ref{eq:DNAw}) and taken from \cite{Cohen2014a}:
\begin{align*}
	\partial_t[\textrm{P}]&=\alpha_{\rm{P}}\frac{w_{\textrm{P}}}
	{w_{\rm{P}}+(1+w_{\textrm{PO}}[\textrm{O}])^2(1+w_{\rm{PN}}[\textrm{N}])^2}-\beta_{\rm{P}}[\textrm{P}]\\
	\partial_t[\textrm{O}]&=\alpha_{\rm{O}}\frac{w_{\rm{O}}(1+w_{\rm{OS}}[\textrm{S}])}
	{w_{\rm{O}}(1+w_{\rm{OS}}[\textrm{S}])+(1+w_{\rm{ON}}[\textrm{N}])^2(1+w_{\rm{OI}}[\textrm{I}])^2}-\beta_{\rm{O}}[\textrm{O}]\\
	\partial_t[\textrm{N}]&=\alpha_{\rm{N}}\frac{w_{\rm{N}}(1+w_{\rm{NS}}[\textrm{S}])}
	{w_{\rm{N}}(1+w_{\rm{NS}}[\textrm{S}])+(1+w_{\rm{NP}}[\textrm{P}])^2(1+w_{\rm{NO}}[\textrm{O}])^2(1+w_{\rm{NI}}[\textrm{I}])^2}-\beta_{\rm{N}}[\textrm{N}]\\
	\partial_t[\textrm{I}]&=\alpha_{\rm{I}}\frac{w_{\rm{I}}}
	{w_{\rm{I}}+(1+w_{\rm{IO}}[\rm{O}])^2(1+w_{\rm{IN}}[\textrm{N}])^2}-\beta_{\rm{I}}[\textrm{I}]
\end{align*}
Here [S] is the level of Sonic Hedgehog signaling - the net amount of Gli activity, see \cite{Cohen2014a} for details. This is taken to be an exponential gradient, $[{\rm S}]=e^{-s/0.15}$ where $s\in [0,1.2]$ labels the neural tube position. The remaining symbols in brackets indicate the concentration of the TFs. Parameters used are
$\alpha_{\rm{P}}=2$, 
$\alpha_{\rm{O}}=2$, 
$\alpha_{\rm{N}}=2$,
$\alpha_{\rm{I}}=2$,
$\beta_{\rm{P}}=2$,
$\beta_{\rm{O}}=2$,
$\beta_{\rm{N}}=2$,
$\beta_{\rm{I}}=2$,
$w_{\rm{PO}}=1.9$,
$w_{\rm{PN}}=26.7$,
$w_{\rm{P}}=3.84$,
$w_{\rm{ON}}=60.6$,
$w_{\rm{OI}}=28.4$,
$w_{\rm{OS}}=180$,
$w_{\rm{O}}=38.24$,
$w_{\rm{NP}}=4.8$,
$w_{\rm{NO}}=27.1$,
$w_{\rm{NI}}=47.1$,
$w_{\rm{NS}}=373$,
$w_{\rm{N}}=21.92$,
$w_{\rm{IO}}=58.8$,
$w_{\rm{IN}}=76.2$,
$w_{\rm{I}}=18.72$.

\clearpage

\end{document}